\documentclass[fleqn,usenatbib]{mnras}

\usepackage{newtxtext,newtxmath}
\usepackage{amsmath}
\usepackage[T1]{fontenc}

\DeclareRobustCommand{\VAN}[3]{#2}
\let\VANthebibliography\thebibliography
\def\thebibliography{\DeclareRobustCommand{\VAN}[3]{##3}\VANthebibliography}

\usepackage{graphicx}	% Including figure files
\usepackage{amsmath}	% Advanced maths commands
\usepackage{caption}  % added
\usepackage{subcaption}

\newcommand{\ud}{\mathrm{d}}
\newcommand{\msun}{\,\mathrm{M}_\odot}

\title[Rotating Magnetoconvection]{3D Simulations of Magnetoconvection in a Rapidly Rotating Supernova Progenitor}

\author[Varma \& M\"uller]{
Vishnu Varma$^{1,2}$\thanks{E-mail: 
v.r.vejayan@keele.ac.uk},
Bernhard M\"uller$^{2}$
\\
$^{1}$
{Astrophysics Group, Lennard-Jones Laboratories, Keele University, Keele ST5 5BG, UK}\\
$^{2}$
School of Physics and Astronomy, 10 College Walk, Monash University, Clayton, VIC 3800, Australia\\
}

\date{Accepted XXX. Received YYY; in original form ZZZ}

\pubyear{2021}

\begin{document}
\label{firstpage}
\pagerange{\pageref{firstpage}--\pageref{lastpage}}
\maketitle

\begin{abstract}
We present a first 3D magnetohydrodynamic (MHD) simulation of oxygen, neon and carbon shell burning in a rapidly rotating $16 \msun$ core-collapse supernova progenitor. We also run a purely hydrodynamic simulation for comparison. After $\mathord\approx180\mathrm{s}$ ($\mathord\approx$ 15 and 7 convective turnovers respectively),  the magnetic fields in the oxygen and neon shells achieve saturation at $10^{11}\mathrm{G}$ and $5\times10^{10}\mathrm{G}$. The strong Maxwell stresses become comparable to the radial Reynolds stresses and eventually suppress convection. The suppression of mixing by convection and shear instabilities results in the depletion of fuel at the base of the burning regions, so that the burning shell eventually move outward to cooler regions, thus reducing the energy generation rate. The strong magnetic fields efficiently transport angular momentum outwards, quickly spinning down the rapidly rotating convective oxygen and neon shells and forcing them into rigid rotation. The hydrodynamic model shows complicated redistribution of angular momentum and develops regions of retrograde rotation at the base of the convective shells. We discuss implications of our results for stellar evolution and for the subsequent core-collapse supernova. The rapid redistribution of angular momentum in the MHD model casts some doubt on the possibility of retaining significant core angular momentum for explosions driven by millisecond magnetars. However, findings from multi-D models remain tentative until stellar evolution calculations can provide more consistent rotation profiles and estimates of magnetic field strengths to initialise multi-D simulations without substantial numerical transients. We also stress the need for longer simulations, resolution studies, and an investigation of non-ideal effects.
\end{abstract}


%%%%%%%%%%%%%%%%%%%%%%%%%%%%%%%%%%%%%%%%%%%%%%%%%%

%%%%%%%%%%%%%%%%% BODY OF PAPER %%%%%%%%%%%%%%%%%%

\section{Introduction}
In recent years, multi-dimensional effects during the advanced convective burning stages of massive stars have received significant interest for multiple reasons, and have been studied extensively by means of hydrodynamic simulations. It has been recognised that seed instabilities from convection play an important dynamical role in core-collapse supernova explosions of massive stars
\citep{Couch2013,Muller2015a,Couch_2015,Mueller_2017,bollig_21,vartanyan_22}.  There is also the question of whether
convective boundary mixing by turbulent entrainment and shell mergers may lead to structural changes in the pre-collapse structure of supernova progenitors and compared to current spherically symmetric stellar evolution models \citep[e.g.,][]{young_05,meakin_07,mueller_16c,jones_17,cristini_17,cristini_19,kaiser_20,rizzuti_23} and affect the nucleosynthesis outcomes from massive stars \citep{ritter_18}. Finally, multi-dimensional simulations of late-stage convective burning are starting to shed light on angular momentum transport and magnetic evolution inside massive stars \citep{Yoshida2021,VarmaMuller2021,McNeill2022}, which is particularly relevant for our understanding of neutron star birth spin rates and magnetic fields and hyperenergetic supernova explosions that are probably driven by rotation and magnetic fields.

Three-dimensional simulations of convection during advanced burning stages in massive stars have so far largely disregarded two important aspects of real stars -- rotation and magnetic fields. The effects of rotation had only been touched upon by the seminal work of \citet{kuhlen_03}, and studies in axisymmetry (2D) of \citet{arnett_10,chatzopoulos_16}, while 3D simulations have only started to explore rotation in recent years \citep{McNeill2022, Yoshida2021, Fields2022}. Similarly, magnetic fields during the advanced burning stages have only been considered recently by \citet{VarmaMuller2021}, and their work was limited to the non-rotating case.

Outside the context of advanced burning stages in massive stars, magnetohydrodynamic (MHD) simulations  have been used extensively as a means to study convection over the years, primarily in the context of the Sun and solar-like stars \citep[for reviews see, e.g.,][]{brun_17,charbonneau_20}. Given the high quality of spatially and temporally resolved solar data \citep[e.g.,][]{Garcia2009, Raouafi2023}, these simulations often aim to explain more detailed observational time-varying features on the solar surface \citep{Rincon2018, Toriumi2019, Petrovay2020} and envelope convection, which includes understanding the formation of the solar rotation profile \citep[e.g.][]{Featherstone2015, Brun2017, Hotta2022, Camisassa2022}.
Studies of stars more massive than the sun are currently limited to just a handful of core dynamo simulations of A and B-type stars \citep[e.g.,][]{Brun2005, Featherstone2009, Auguston2016}. Recently, 3D simulations \citep{Raynaud2020, Masada2022} have also shown that the convective dynamo can be an important ingredient in amplifying magnetic fields in newly forming neutron stars.

Simulations of magnetoconvection during the late burning stages, both in rotating and non-rotating stars, are a necessity for several reasons. Even in slowly rotating massive stars, magnetic fields have been shown to impact the dynamics of the subsequent neutrino-driven explosions \citep{obergaulinger_14,MullerVarma2020, Matsumoto2020, Matsumoto2022, Varma2022}. For the magnetorotational explosion scenario  (e.g., \citealp{burrows_07a,Winteler2012a,Mosta2014,moesta_18,Obergaulinger2020,Obergaulinger2020b,kuroda_20,aloy_21,powell_23}; see also early work on explosions driven by millisecond magnetars, e.g., \citealp{Usov1992, Duncan1992}), a better understanding of the interplay between convection, rotation, and magnetic fields in supernova progenitors is even more critical. In this mechanism, a rapidly rotating core and very strong initial magnetic fields are required to launch the very energetic explosion. Such magnetorotational explosions are thought to explain rare, unusually energetic ``hypernovae'' with energies of up to $\mathord{\sim}10^{52}\,\mathrm{erg}$ \citep{woosley_06}. 

The magnetorotational explosion mechanism is linked to the problem of rotation and magnetism in massive stars. Initial conditions for magnetorotational explosion simulations currently come from ``1.5D'' stellar evolution models that assume shellular rotation and include effective recipes for magnetic field generation and angular momentum transport by hydrodynamic and magnetohydrodynamic processes \citep{Heger2000, Heger2005, woosley_06}. 

There are still many open questions about the treatment of rotation and magnetic fields in stellar evolution models.
Aside from purely hydrodynamic instabilities \citep{Heger2000, Heger2005, Maeder2013}, the interaction of rotation and magnetic fields is a critical issue. Since convective regions are usually assumed to rotate rigidly (as this is the only allowed rotational state in thermal equilibrium; \citealp{Landau1969}), attention has usually focused on angular momentum transport and dynamos in non-convective regions.
The dynamo mechanism often implemented in these 1.5D stellar models to generate magnetic fields relies on sufficiently strong differential rotation in convectively stable regions of the star to stretch poloidal magnetic fields into toroidal fields. The dynamo loop is closed by the development of a pinch-type (Pitts-Tayler) instability \citep{Tayler1973, Pitts1985}. This mechanism developed by \citet{Spruit1999, Spruit2002} is often referred to as the Tayler-Spruit dynamo.
Recently, \citet{fuller_19} have tried to improve the Tayler-Spruit dynamo mechanism, arguing that the Tayler instability saturates via turbulent dissipation of unstable magnetic field perturbations. This mechanism has a smaller energy dissipation rate and thus allows for stronger magnetic fields and more efficient angular momentum transport than the traditional Tayler-Spruit dynamo. 

Other attempts to understand magnetic stellar evolution models have been to derive scaling relationships in convective regions \citep{Jermyn2020, Jermyn2020b} and to explore the role of the magnetorotational instability (MRI) \cite{Wheeler2015, takahashi_20}, driven, in part, by global 3D simulations such as that in \cite{Braithwaite2006, Jouve2015, Meduri2019}. These simulations have suggested that the interaction between the different instabilities and flows can be quite intricate, and may induce not only the pinch instability but can also be strongly affected by MRI and magnetic buoyancy. 

Since 1.5D stellar evolution models implementing the Tayler-Spruit dynamo predict magnetic fields that are rather weak and predominantly toroidal, the general notion has long been that field amplification processes after the collapse are critical in magnetorotational explosions \citep{Obergaulinger2009, Sawai2013, Sawai2015, Masada2015, Guilet2015a, Guilet2015b, Moesta2015, Rembiasz2016, Rembiasz2016b, Raynaud2020, Reboul2021}, although this has recently been challenged \citep{Obergaulinger2017, Obergaulinger2021, aloy_21}. In particular, for sufficiently strong seed fields in the progenitor, the initial
field strengths and geometry could have a significant
impact on the development of magnetorotational explosions after collapse \citep{bugli_20,aloy_21}, making an understanding of the pre-collapse magnetic fields in 3D indispensable. 

In this study, we present a first simulation of rotating magnetoconvection
during the final phases of shell burning using the
ideal MHD approximation. 
This simulation constitutes a first step beyond spherically symmetric prescriptions in stellar evolution models
to predict the magnetic field strength and geometry, as well as its role in angular momentum transport 
encountered in the inner shells of massive stars at the pre-supernova
stage. We also compare to a corresponding non-magnetic model of the same progenitor to gauge the feedback of magnetic fields on the convective flow and rotation profiles.

Our paper is structured as follows. In Section \ref{sec:methods}, we describe the
numerical methods, progenitor model, and initial conditions
used in our study. The results of the simulations are presented in Section~\ref{sec:results}. We first focus on the strength and geometry of the emerging magnetic field and then analyse the impact of magnetic fields on the convective flows and rotation, with a focus on the turbulent mixing and angular momentum transport within and between the burning shells. We summarise our results and discuss their implications in Section~\ref{sec:conclusion}. 

\section{Numerical Methods and Simulation Setup}
\label{sec:methods} 

\begin{figure}
 \includegraphics[width=\columnwidth]{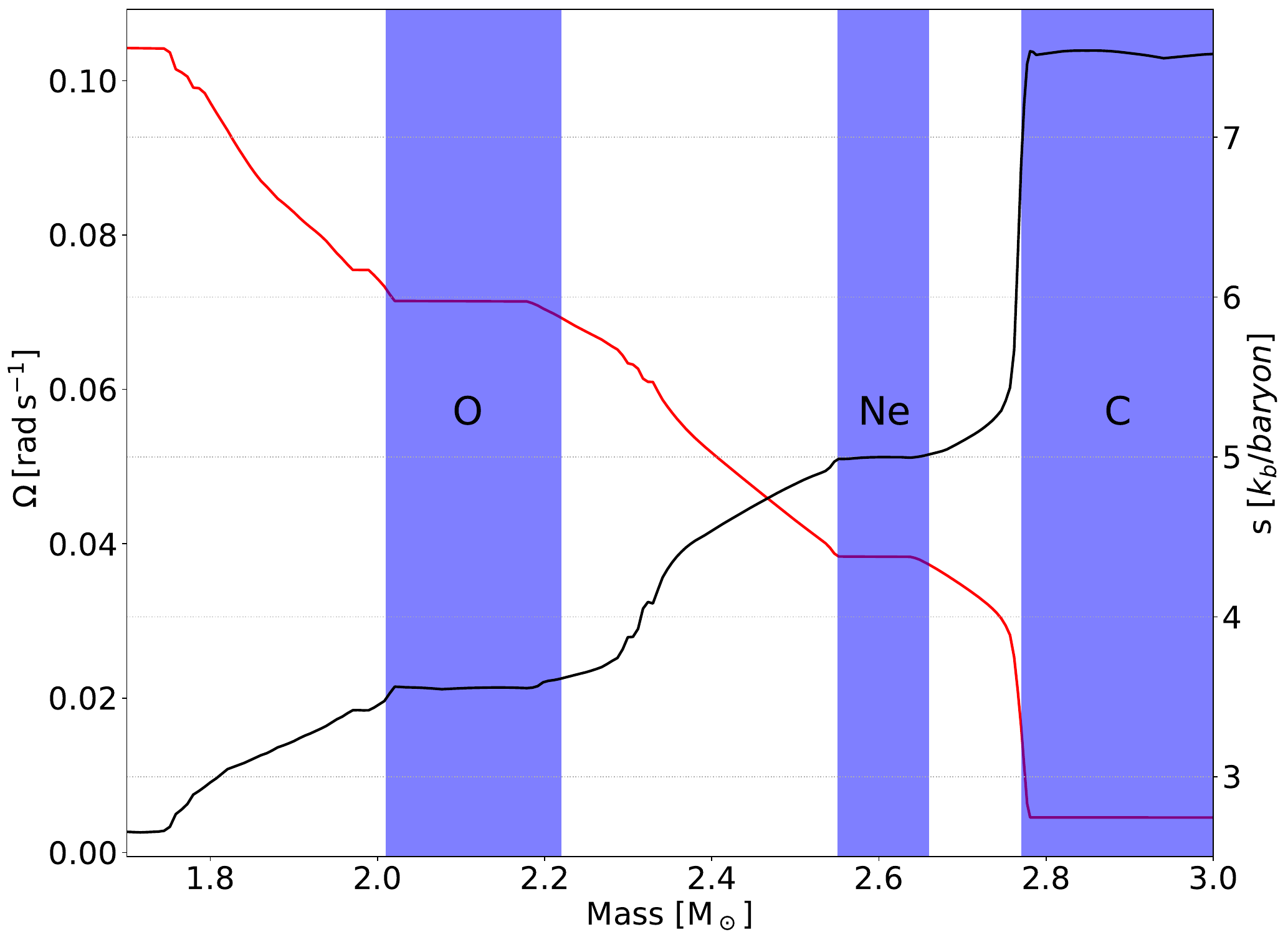}
 \caption{\normalsize Initial profiles of rotation rate, $\Omega$ (red), and density, $\rho$ (black), of the $16 \msun$ progenitor when mapped to \textsc{CoCoNuT}. Note that the figure starts at $\mathbf{1.7\mathrm{\msun}}$ ($\mathbf{3,000\,\mathrm{km}}$), which is the inner domain boundary of the simulation, until $\mathbf{3.0\mathrm{\msun}}$  ($\mathbf{12,000\, \mathrm{km}}$) to easily see the important features of the profile. The simulated domain extends to $\mathbf{6.2\mathrm{\msun}}$ ($\mathbf{40,000\mathrm{km}}$). The shaded bands depict the initial positions of the oxygen, neon, and carbon convective burning shells.}
  \label{fig:Initial_Entropy}
\end{figure}

We simulate oxygen, neon, and carbon shell burning 
with and without magnetic fields
in a rapidly rotating $16 \msun$ solar-metallicity helium star from \citet{woosley_06} with a strong differential rotation profile calculated using the  stellar evolution code \textsc{Kepler}. The same progenitor model has previously been used in the \textsc{PROMETHEUS} rotating shell convection simulation of \citet{McNeill2022}. 
The structure of the stellar evolution model at
the time of mapping to 3D is illustrated in Figure~\ref{fig:Initial_Entropy}. 
% The model contains
% two active convective shells with sufficiently
% short turnover times to make time-explicit simulations feasible.
% The oxygen shell extends from 
% $1.75\, \msun$ to $2.25\, \msun$ in enclosed
% mass and from $3,400\, \mathrm{km}$
% to $7,900\,\mathrm{km}$ in radius, immediately
% followed further out by the neon shell
% out to $3.53\, \msun$ in mass and
% $27,000\, \mathrm{km}$ in radius.

For our 3D simulations we employ the Newtonian magnetohydrodynamic (MHD) version of the  \textsc{CoCoNuT} code as described in \citet{MullerVarma2020, VarmaMuller2021}.  
The MHD equations are solved in spherical
polar coordinates using the HLLC (Harten-Lax-van Leer-Contact) Riemann solver \citep{Gurski2004a, Miyoshi2005a}. The divergence-free condition $\nabla\cdot\mathbf{B} = 0$ is maintained using a modification
of the original hyperbolic divergence cleaning scheme of \citet{Dedner2002} that allows for a variable cleaning speed while still maintaining total energy conservation as described in \citet{VarmaMuller2021}
(building on similar ideas by \citealt{Tricco2016}).
The extended system of MHD equations for the density $\rho$, velocity $\mathbf{v}$, magnetic field $\mathbf{B}$,  total energy density $\hat{e}$, mass
fractions $X_i$, and the rescaled Lagrange multiplier $\hat{\psi}$ reads,
\begin{eqnarray}
\partial_t \rho
+\nabla \cdot \rho \mathbf{v}
&=&
0,
\\
\partial_t (\rho \mathbf v)
+\nabla \cdot \left(\rho \mathbf{v}\mathbf{v}-
\frac{\mathbf{B} \mathbf{B}}{4\pi}
+P_\mathrm{t}\mathcal{I}
\right)
&=&
\rho \mathbf{g}
-
\frac{(\nabla \cdot\mathbf{B}) \mathbf{B}}{4\pi}
,
\\
\partial_t {\hat{e}}+
\nabla \cdot 
\left[(e+P_\mathrm{t})\mathbf{u}
-\frac{\mathbf{B} (\mathbf{v}\cdot\mathbf{B})
{-c_\mathrm{h} \hat{\psi} \mathbf{B}}}{4\pi}
\right]
&=&
\rho \mathbf{g}\cdot \mathbf{v}
+
\rho \dot{\epsilon}_\mathrm{nuc}
,
\\
\partial_t \mathbf{B} +\nabla \cdot (\mathbf{v}\mathbf{B}-\mathbf{B}\mathbf{v})
+\nabla  (c_\mathrm{h} \hat{\psi})
&=&0,
\\
\partial_t \hat{\psi}
+c_\mathrm{h} \nabla \cdot \mathbf{B}
&=&-\hat{\psi}/\tau,
\\
\partial_t (\rho X_i)
+\nabla \cdot (\rho  X_i \mathbf{v})
&=&
\rho \dot{X}_i
,
\end{eqnarray}
where $\mathbf{g}$ is the gravitational acceleration, $P_\mathrm{t}$ is the total (gas and magnetic) pressure, $\mathcal{I}$ is the Kronecker tensor, $c_\mathrm{h}$ is the hyperbolic cleaning speed, $\tau$ is the damping time scale for divergence cleaning, and $\dot{\epsilon}_\mathrm{nuc}$ and $\dot{X}_i$ are energy and mass fraction
source terms from nuclear reactions. This
system conserves the volume integral of a
modified total energy density $\hat{e}$,
which also contains the cleaning field $\hat{\psi}$,
\begin{equation}
{\hat{e}}
=\rho \left(\epsilon+\frac{v^2}{2}\right)+\frac{B^2+\hat{\psi}^2}{8\pi},
\end{equation}
where $\epsilon$ is the mass-specific internal energy.

The simulations are conducted on a grid
with $400\times128\times256$ zones in radius $r$, colatitude $\theta$, and longitude $\varphi$
with an exponential grid in $r$ and uniform
spacing in $\theta$ and $\varphi$.
To reduce computational costs, we excise the non-convective inner core up to $3,000\,\mathrm{km}$ and replace the excised core with a point mass. The grid extends to a radius
of $40,000\, \mathrm{km}$ and includes a small part
of the silicon shell, the entire convective oxygen, neon and carbon shells. Our simulations cover the full sphere ($4\pi$ solid angle).

In the MHD simulation, we impose a homogeneous magnetic field with $B_z = 10^{7}\,\mathrm{G}$ parallel to the grid axis as initial conditions. 

We implement reflecting and periodic boundary conditions in $\theta$ and $\varphi$, respectively. For the hydrodynamic variables, we use hydrostatic
extrapolation \citep{mueller_20review} at the inner and outer boundary, and
impose an effectively slip-free inner boundary. 
Different from the hydrodynamic simulations of \citet{Yoshida2021} and \citet{McNeill2022}, we do \emph{not} contract
the inner boundary to follow the contraction and
collapse of the core. The model is instead meant to be representative of the physical principles governing late-stage, steady-state magnetoconvection with rapid rotation, rather than a pre-collapse model.

The inner and outer boundary conditions for the
magnetic fields are less trivial.
In simulations of magnetoconvection in the Sun, 
various choices such as vertical boundary conditions
($B_x=B_y=0$), radial boundary conditions ($B_\theta=B_\varphi=0$), vanishing tangential
electric fields or currents, perfect-conductor
boundary conditions, or extrapolation
to a potential solution have been employed
\citep[e.g.,][]{Thelen2000, Rempel2014,Kapyla2018}. 
Since our domain boundaries are separated from
the convective regions by shell interfaces with
significant buoyancy jumps, we opt for the
simplest choice of boundary conditions and merely
fix the magnetic fields in the ghost zones
to their initial values for a homogeneous vertical magnetic field. We argue that due to the buffer regions at our radial boundaries, and the lack of rotational shear (due to the slip-free boundary conditions), our choice of magnetic boundary conditions should not have a significant impact on the dynamically relevant regions of the star. 

Similar to the non-rotating magnetoconvection simulations done in \citet{VarmaMuller2021}, our models will not (and are not
intended to) provide an exact representation of the
pre-collapse state of the particular $16 \msun$ star
that we are simulating. We would expect, e.g., that
for the particular $16 \msun$ model, the burning rate
and hence the convective velocities would increase
until the onset of collapse due
to the contraction of the convective oxygen shell. As a consequence
of accelerating convection and flux compression, the magnetic
fields will likely also be somewhat higher at the onset of collapse.
The model is rather meant to reveal the physical principles
governing late-stage magnetoconvection in rapidly rotating massive stars,
and to be \emph{representative} of the
typical conditions in the burning shells with the understanding
that there are significant variations in convective Mach
number and shell geometry at the onset of collapse \citep{Collins2018},
which will also be reflected in the magnetic field strengths in the
interiors of magnetorotational supernova progenitors.

\section{Results}
\label{sec:results}

\subsection{Evolution of the magnetic fields}
\label{subsec:Magnetic}

\begin{figure}
 \includegraphics[width=\columnwidth]{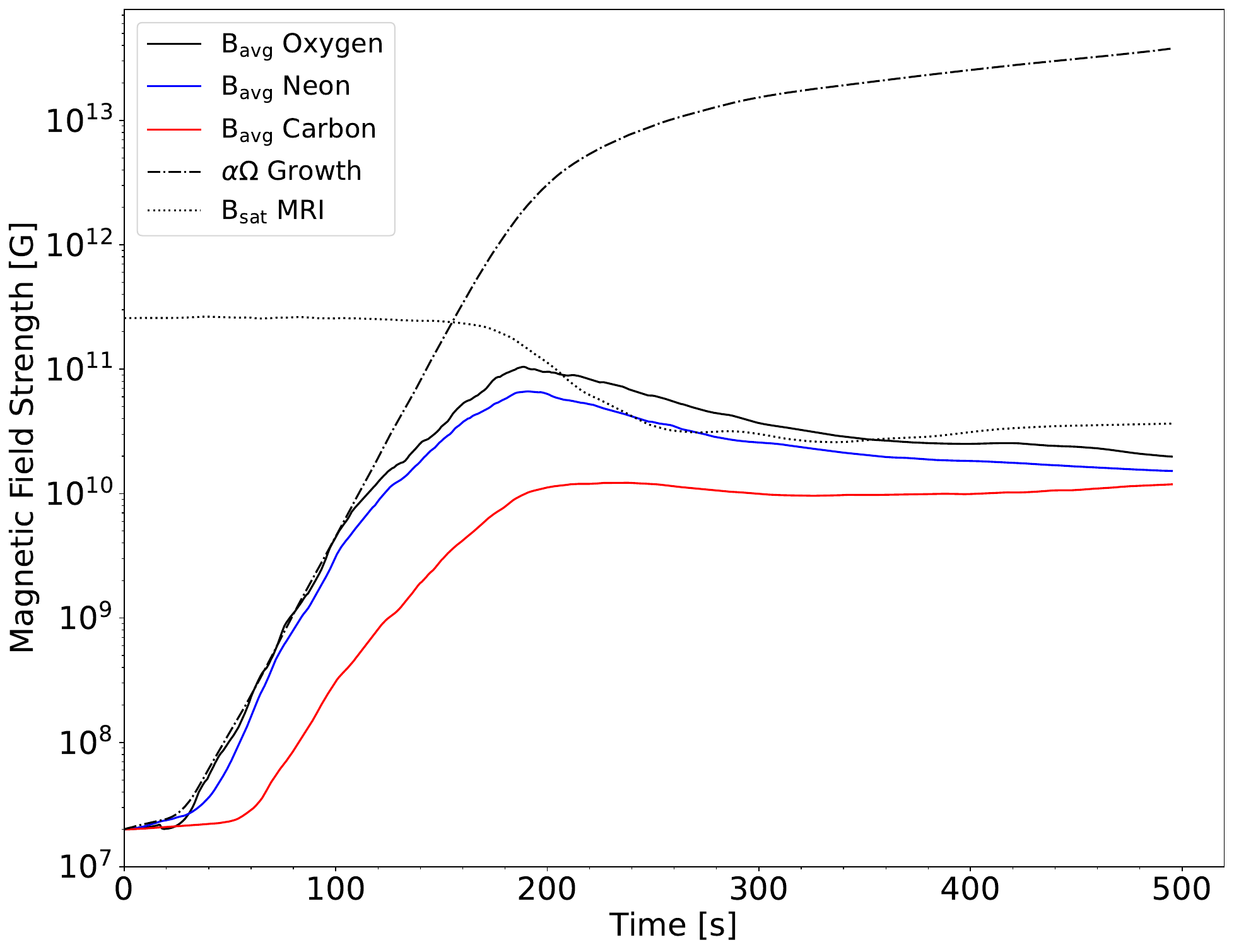}
 \caption{\normalsize Evolution of the volume averaged (solid) magnetic field strength within the oxygen (black), neon (blue) and carbon (red) shells. The dashed line approximates the expected growth of the magnetic fields strength in the  oxygen shell via the $\alpha\Omega$ dynamo mechanism described in Section \ref{subsec:Magnetic}. The dotted line shows the expected magnetic field strength saturation due to differential rotation
 in the oxygen shell based on Equation \ref{eq:B_sat}.}
 \label{fig:BEvolve}  
\end{figure}

We simulate two rapidly rotating $16 \msun$  models, one with and one without magnetic fields. The magnetic model is initiated with a homogeneous magnetic field of $10^7 \mathrm{G}$. We then allow the geometry of the magnetic field to evolve naturally under the influence of rapid rotation and convection. 

The evolution of the root mean square (RMS), volume-averaged magnetic fields in the three convective burning shells we simulate ---
the oxygen, neon and carbon burning shell ---
are shown in Figure~\ref{fig:BEvolve}. We see an initial period of exponential growth of magnetic field strength in each shell before a plateau forms after $\mathord\approx200\mathrm{s}$ in each shell. The field strengths in the oxygen and neon shells both appear to follow a very similar trajectory, achieving a peak at $\mathord\approx190\mathrm{s}$, before a gradual decline sets in.
In each of the shells, convection takes a different amount of time to fully develop, which explains the slight delay from the start of the simulation to the beginning of the exponential field growth. In particular, the carbon shell has a much longer convective turnover timescale $\tau_\mathrm{c}$ than the other two shells
with an initial value
$\tau_\mathrm{c}\approx300\, \mathrm{s}$ compared to $\mathord\approx 15\, \mathrm{s}$ and $25\,\mathrm{s}$ for the oxygen and neon shell, respectively, which considerably delays the growth of  magnetic fields. The growth of the magnetic fields in the carbon shell already becomes apparent after $\mathord\approx 60\,\mathrm{s}$, even without convection being fully developed. This is due to field amplification from strong differential rotation which develops at the base of the shell, and turbulent fluctuations that developed alongside the convective plumes.

Due to the development of convection in the shells, coupled with rapid differential rotation (maximum rotation rate of $\mathrm{\Omega \, \approx\,0.104\,rad\,s^{-1}}$), we expect the field amplification to be dominated by the $\alpha\Omega$ dynamo mechanism, which is often proposed as the mechanism that sustains the solar magnetic field \citep{charbonneau_20}. 
The mechanism stretches the poloidal magnetic fields into toroidal fields via differential rotation ($\Omega$-mechanism), and the toroidal field is then stretched into a poloidal field due to convective motions ($\alpha$-mechanism), completing the cycle and amplifying the seed field.
To test if this is the case, we plot the expected growth of the $\alpha\Omega$ dynamo for the oxygen shell in Figure \ref{fig:BEvolve}.

To this end, we approximate the growth of the magnetic field via the $\alpha\Omega$ mechanism in the oxygen shell by assuming the magnetic field evolves via the simplified evolution equation:
\begin{equation}
    \frac{{\partial}B_\mathrm{rms}}{{\partial}t} = \Gamma_{\alpha\Omega}B_\mathrm{rms},
\end{equation}
\label{eq:Bev}
which has a solution for the magnetic field growth of the form $B_\mathrm{rms} = B_0e^{\Gamma_{\alpha\Omega}\Delta t}$, where $B_0$ is the initial field strength. We take the growth rate of the $\alpha\Omega$ dynamo to be $\Gamma_{\alpha\Omega} = (v/L)(\Omega\tau_\mathrm{c})^{1/2}$ as presented in \citet{Vishniac2005} based on dimensional arguments, where $v$ is the convective velocity, $L$ is the radial extent of the convective zone, $\Omega$ the rotation rate and $\tau_c$ is the convective turnover timescale.
Since $\tau_\mathrm{c}\sim L/V$, this effectively amounts to a growth time scale $\tau_{\alpha\Omega}$ of the order of the geometric mean of the rotation period $P=2\pi \Omega^{-1}$ and the convective turnover time, $\tau_{\alpha\Omega}=\Gamma_{\alpha\Omega}^{-1}\sim (2 \pi)^{-1/2}\sqrt{P \tau_\mathrm{c}}$.
In the evolution plotted in Figure \ref{fig:BEvolve}, we first calculate the RMS averaged values $v$ and $\Omega$ in the oxygen shell, as well as angular averages the radial extent of the oxygen shell, $L$. The convective turnover time is calculated using these averaged quantities, $\tau_\mathrm{c}\, = L/v$. These averaged quantities are used to then determine the growth rate, $\Gamma_{\alpha\Omega}$ to be used at each time step, to evolve the magnetic field.

The expected growth rate of the $\alpha\Omega$ dynamo follows the growth of the magnetic field in the oxygen shell very closely for the first $\mathord{\approx }140\,\mathrm{s}$, after which time the magnetic field growth in the simulation slows down, and eventually decays after hitting a peak magnetic field strength of $\mathord{\approx} \, 10^{11}\,\mathrm{G}$ and $7\times10^{10}\,\mathrm{G}$ for the oxygen and neon burning shells, respectively. We see that the expected growth rate from an $\alpha\Omega$ dynamo also decreases at later times due to two factors. First, convective velocities drop due to suppression of convection by strong magnetic stresses. Second the rotation rate drops due to large angular momentum fluxes. We will discuss these effects further in Sections~\ref{subsec:Convection&Energy} and \ref{subsec:Rotation}.

These effects stop the magnetic field from being amplified via the $\alpha\Omega$ dynamo. Since the convection dies down, it
is reasonable to expect that late-time field amplification and saturation is determined by differential rotation alone.  Interestingly, the saturation of the field appears to be well described by an amplification mechanism that is driven by the MRI. For the MRI, \citet{akiyama_03} argue that the saturation field is roughly given by
\begin{equation}
B_\mathrm{sat}^2
\propto 4\pi \rho r^2 {\Omega}^2 \frac{d\ln\Omega}{d\ln r},
\label{eq:B_sat}
\end{equation}
where $\rho$ is the density, $\Omega$ the rotation rate, $r$ the radius and $\frac{d\ln\Omega}{d\ln R}$ quantifies the amount of differential rotation present. 
\citet{akiyama_03} derive
Equation~(\ref{eq:B_sat}) by assuming that saturation of the magnetic field is achieved in the star when the characteristic mode scale 
$l_\mathrm{mode}\approx v_\mathrm{A} 
(\ud\ln\Omega/ \ud \ln r)^{-1}$ is equal to the local radius $r$, since the wavelength of the 
mode cannot be larger than the physical size of the unstable region. Here, $v_\mathrm{A}$ is the Alfv\'en velocity ($v_\mathrm{A} =  B/\sqrt{4\pi\rho}$). On dimensional grounds, one 
can expect Equation~(\ref{eq:B_sat}) to hold not just specifically for the MRI, but more broadly for amplification mechanisms driven solely by differential rotation in the ideal MHD regime with negligible resistivity.
Figure~\ref{fig:BEvolve} shows that the magnetic field in the oxygen shell saturates at a very similar level to Equation~(\ref{eq:B_sat}).

The fastest-growing MRI mode has the wavenumber
\begin{equation}
k_\mathrm{MRI} = \sqrt{1 - \frac{1}{4} \left(2 - \frac{d\ln\Omega}{d\ln r}\right)^2} \frac{\Omega}{c_{A_z}} ,
\label{eq:k_mri}
\end{equation}

where $c_{A_z} = b_{0_z}/\sqrt{\rho}$ is the Alfv\'en velocity in the vertical direction \citep{Balbus1991, Rembiasz2016}. In our simulation, this corresponds to a wavelength of $\approx 10^6\mathrm{cm}$ at early times when the field growth is dominated by the $\alpha\Omega$ dynamo. After the field saturates, and we expect differential rotation to play a key role in maintaining the magnetic fields, we find $k_\mathrm{MRI}\approx 10^8
\,\mathrm{cm}$, which is well resolved in our simulated grid($\mathrm{dr} \leq 6\times 10^6 \,\mathrm{cm}$ within the oxygen and neon shells). Although buoyancy forces can impact the MRI modes \citep{Obergaulinger2009}, the regions we are considering are well-mixed with a flat entropy gradient, so it is justified to ignore buoyancy.

The strong magnetic fields also result in very rapid redistribution of angular momentum, which slows the rotation rate of the oxygen and neon shells dramatically. Since the magnetic field saturation depends on the rate of rotation, this in turn leads to a drop in the average magnetic field strength by over 50\% in these shells, as we see in Figure~\ref{fig:BEvolve}. The consequences of this will be discussed in more detail in the next sections. 

\begin{figure}
 \includegraphics[width=\columnwidth]{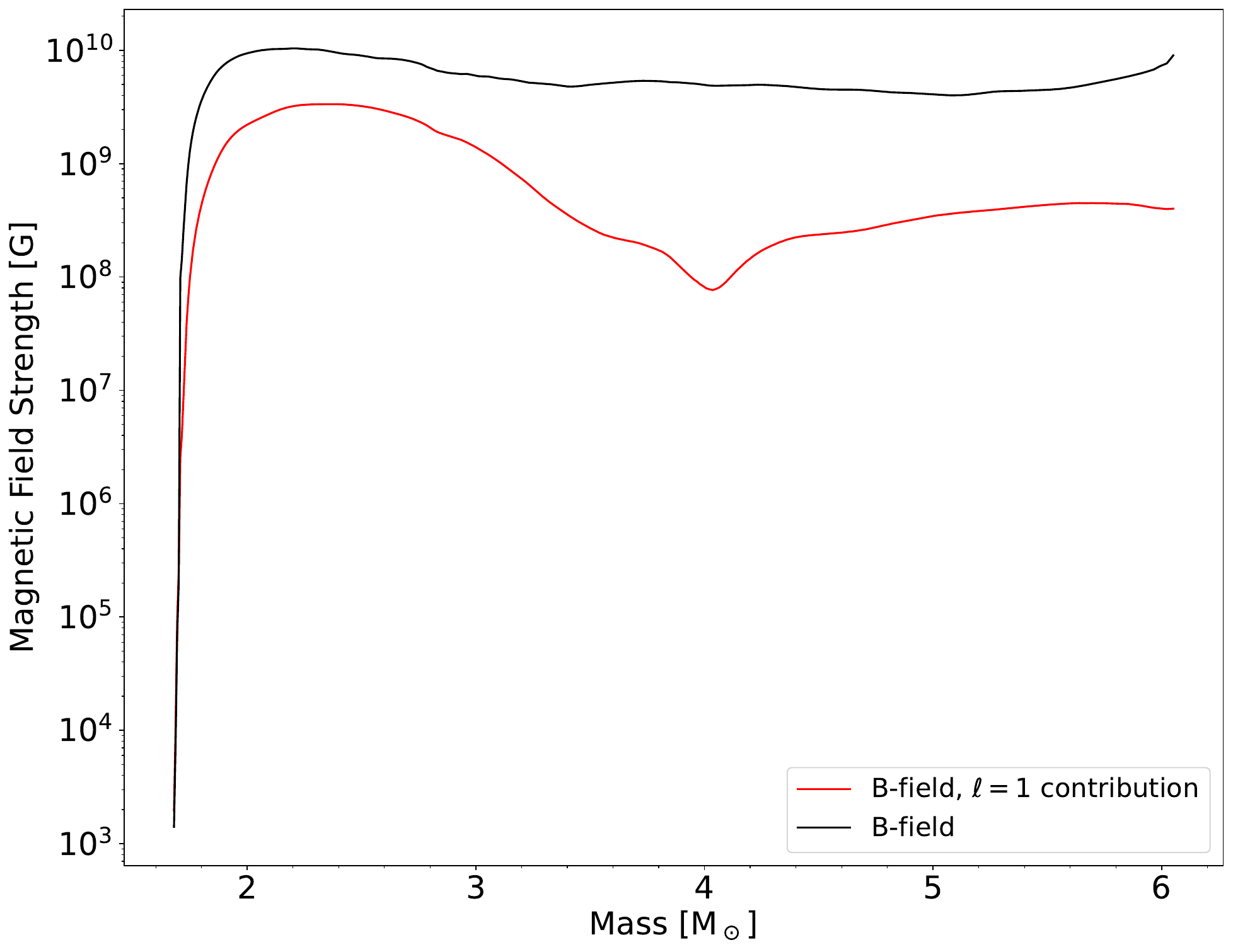}
 \caption{\normalsize The angle-averaged RMS value (black) and
the dipole component (red) of the radial magnetic field component as a function of mass coordinate at a time of $480\, \mathrm{s}$.}
  \label{fig:Dipole}  
\end{figure}

\begin{figure*}
\centering
    \subfloat[]{\includegraphics[width=0.32\linewidth]{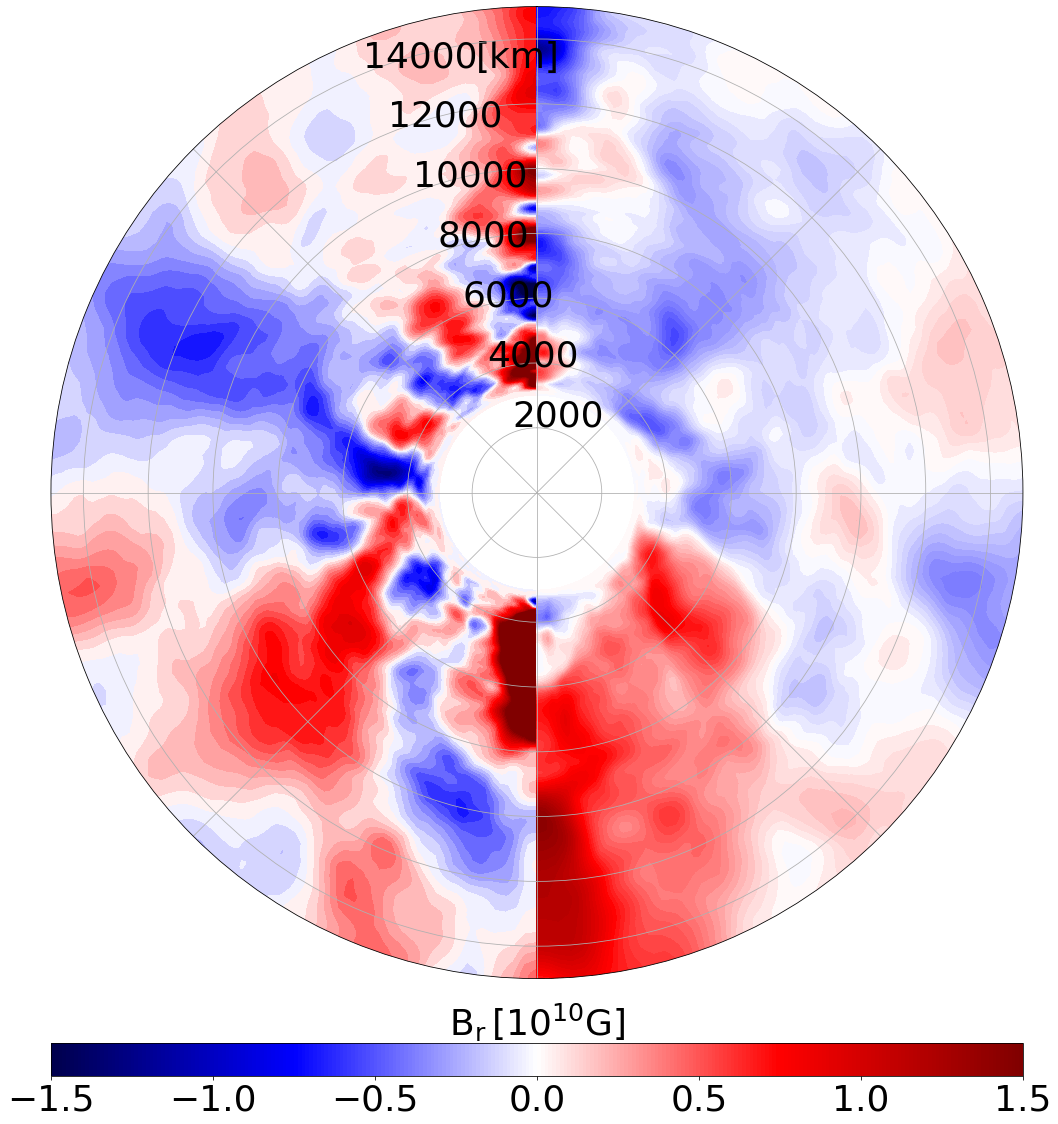}}
\hfil
    \subfloat[]{\includegraphics[width=0.32\linewidth]{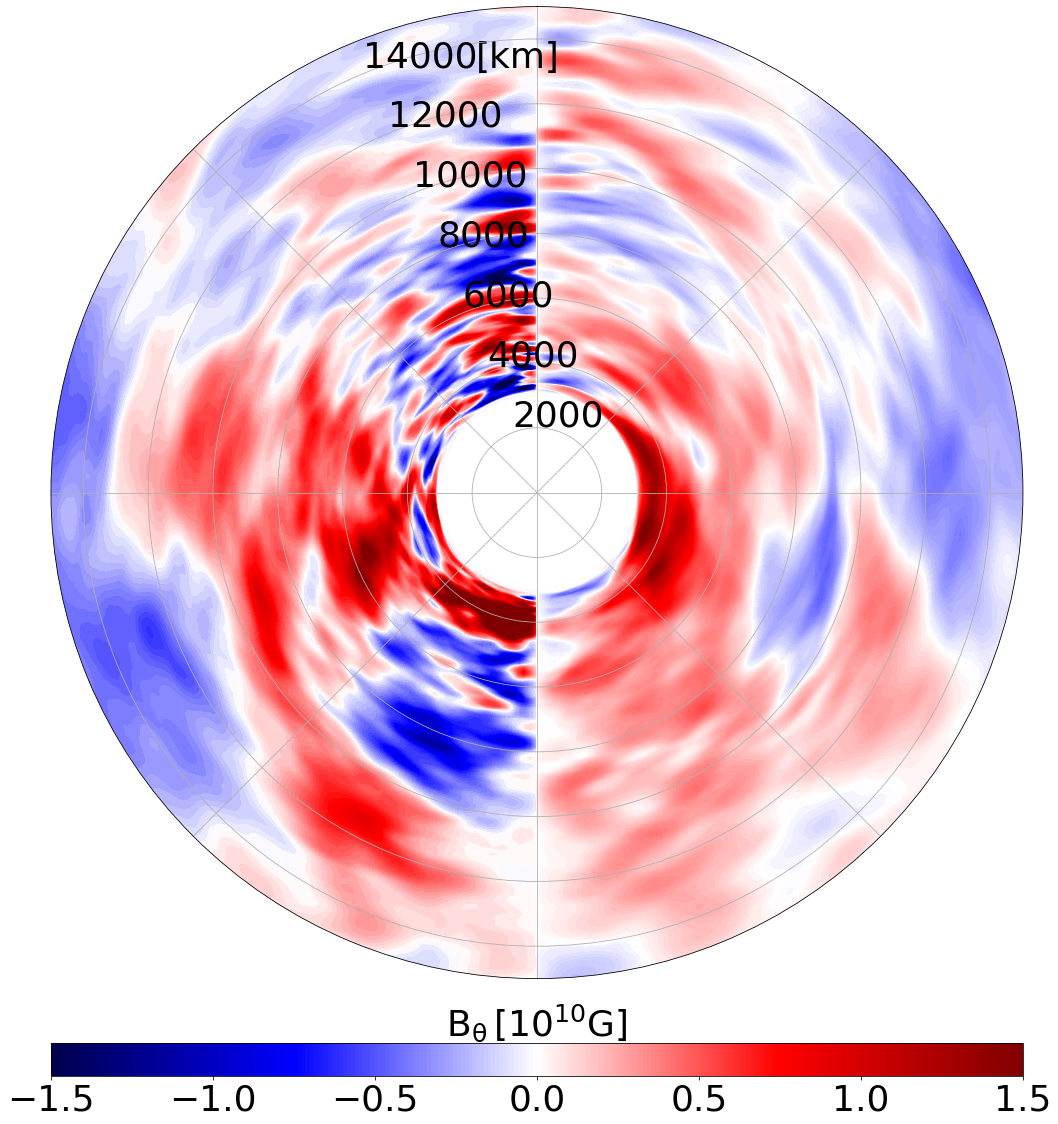}}
\hfil
    \subfloat[]{\includegraphics[width=0.32\linewidth]{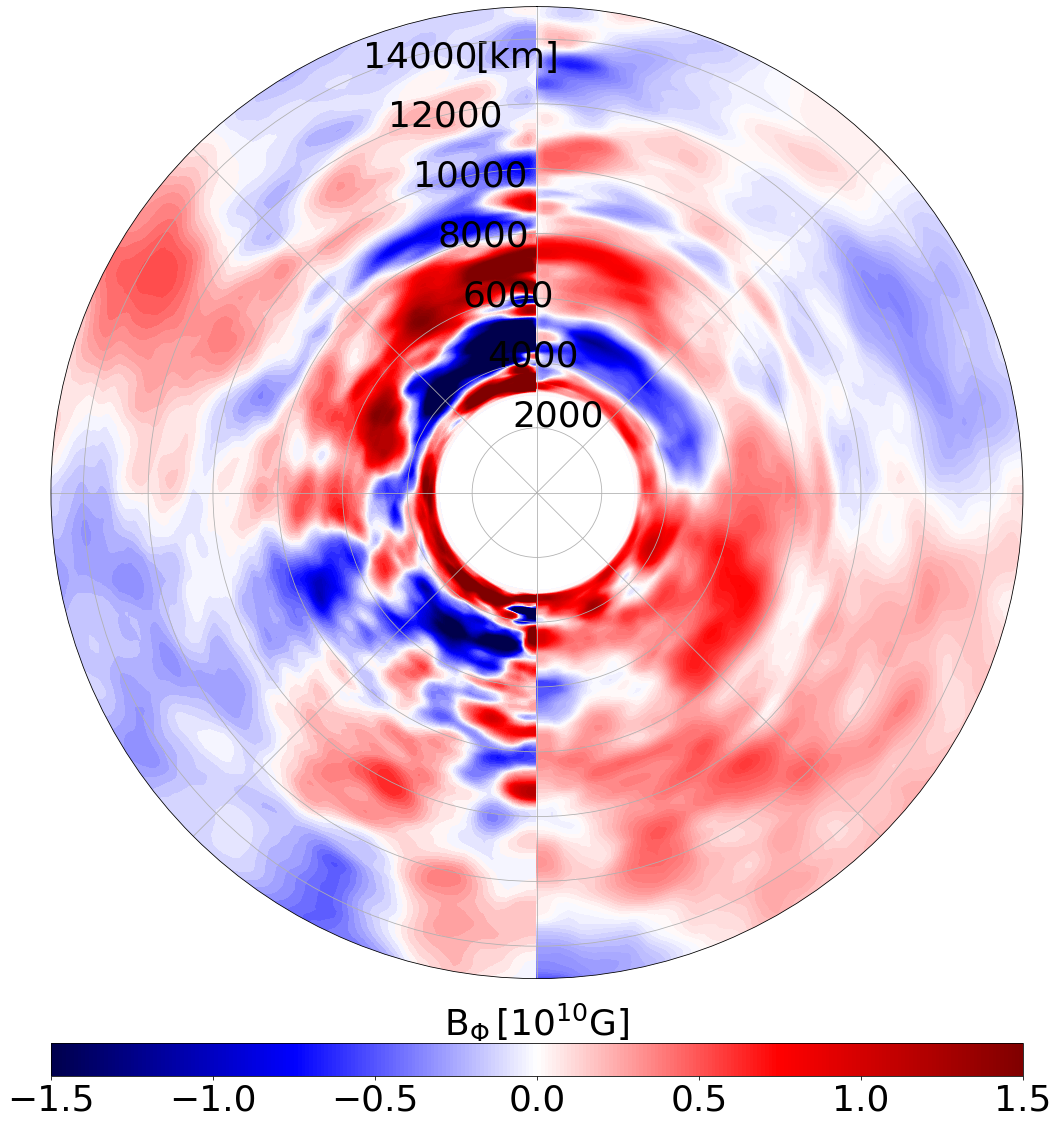}}

\caption{\normalsize Zonal averages of the magnetic field components, $\mathrm{B}_r$, $\mathrm{B}_
\theta$ and $\mathrm{B}_\phi$, shown in panels (a), (b) and (c) respectively. The left halves of each image are presented at a time of $\approx 250\,\mathrm{s}$, and the right halves at $\approx 480\mathrm{s}\, $. The averages are plotted up to a radial coordinate of $\mathrm{15,000\,km}$, which includes the entire oxygen and neon shells, and the innermost part of the carbon shell.}
    \label{fig:Geometry}
\end{figure*}

\begin{figure*}
\centering
    \subfloat[]{\includegraphics[width=0.45\linewidth]{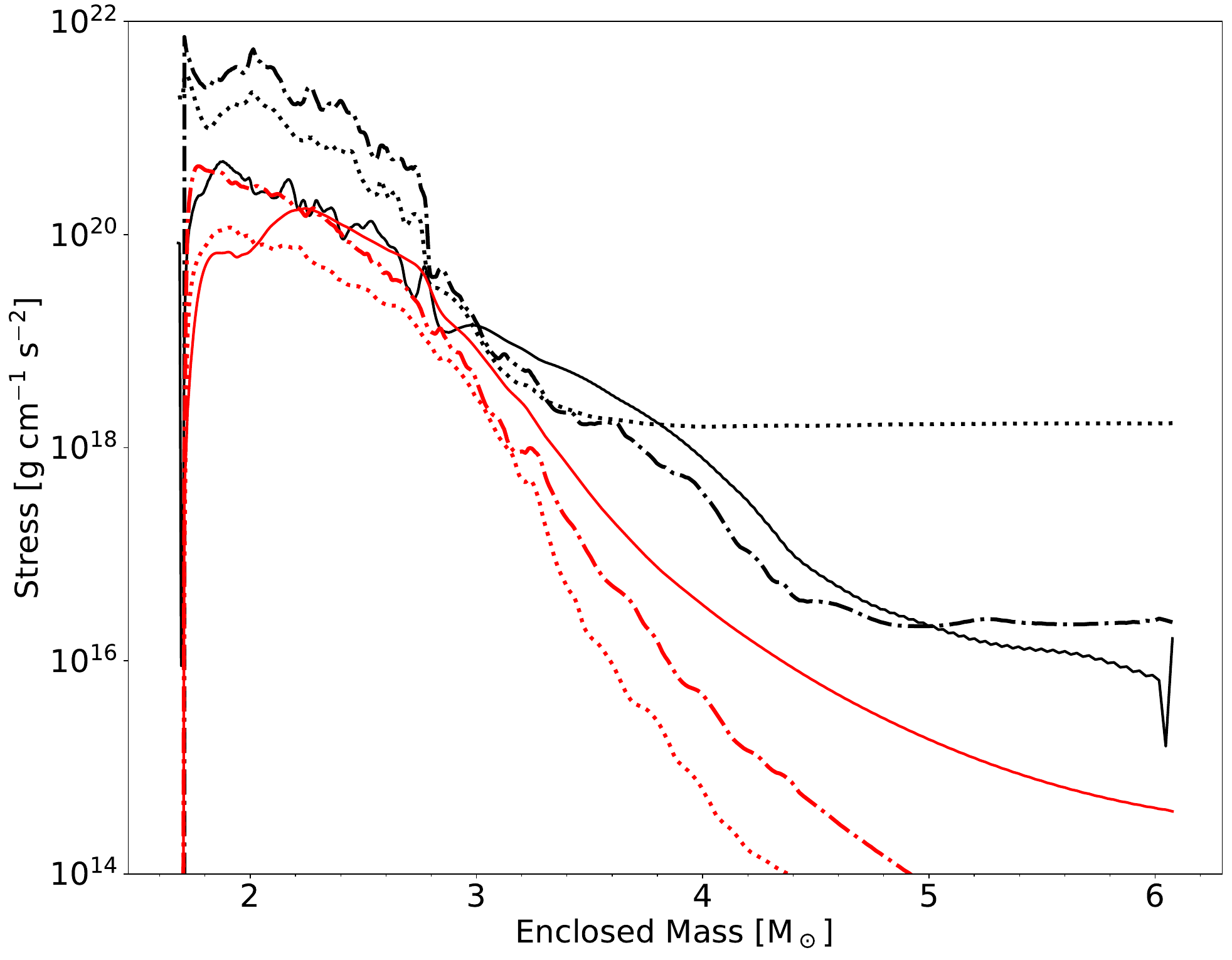}}
\hfil
    \subfloat[]{\includegraphics[width=0.45\linewidth]{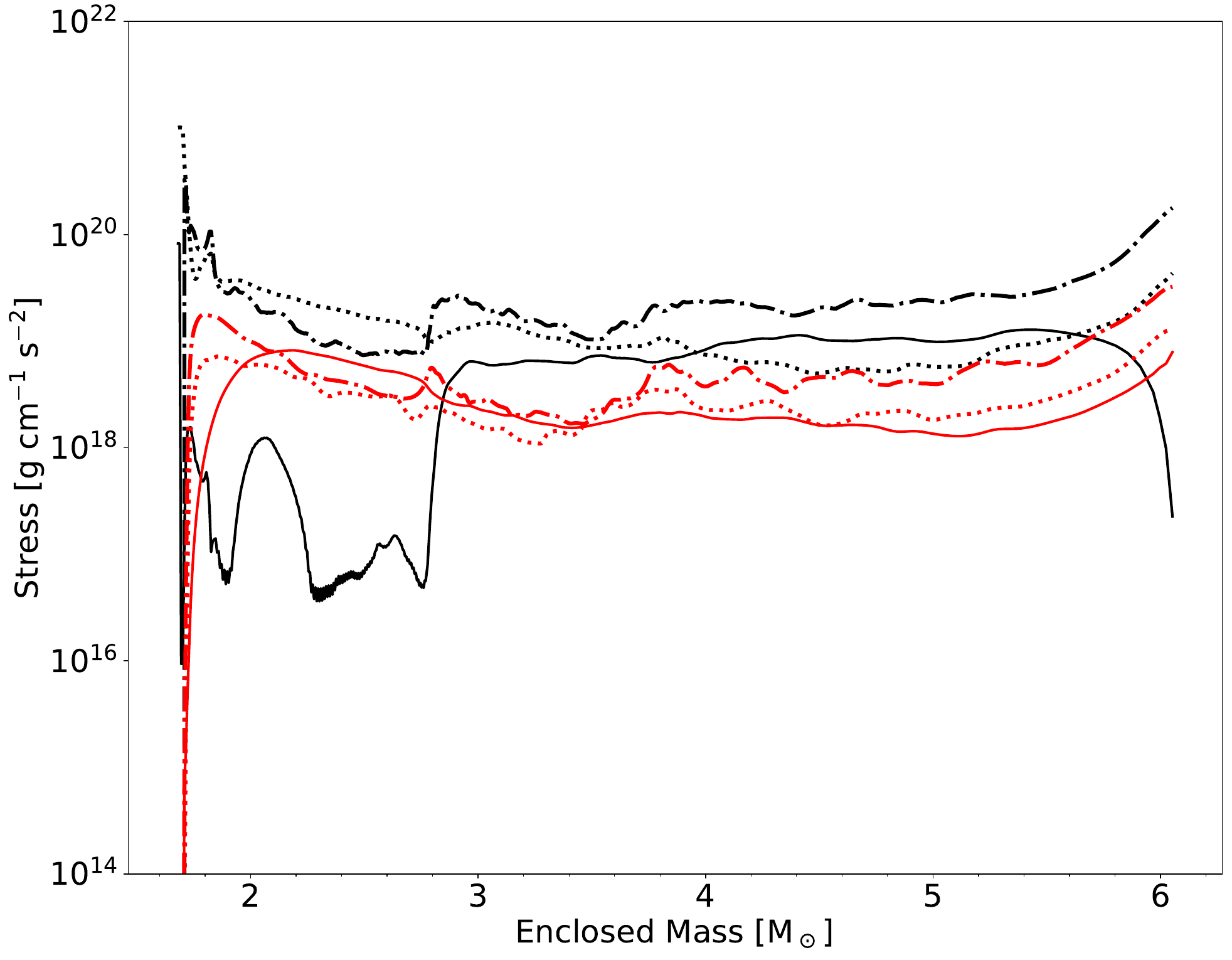}}

\caption{\normalsize The radial  (solid) and non-radial (dashed)
 diagonal components of the 
 Reynolds stress tensor
 $R_{ij}$ (black) and  Maxwell 
 tensor $M_{ij}$ (red) 
 for the MHD convection model at $180\,\mathrm{s}$ (a) and $480\,\mathrm{s}$ (b) as a function of enclosed mass. The stress tensors are defined in Equation~ (\ref{eqn:Stress1},\ref{eqn:Stress2}). }
    \label{fig:Stress}
\end{figure*}

Aside from the equilibrium field strength, it is worth investigating the field geometry that has naturally developed in the saturation state.
To this end, we show radial profiles of the
RMS averaged field strength
and of the dipole field strength in Figure~\ref{fig:Dipole}, as well as zonal averaged plots of $\mathrm{B_r, B_{\Theta}}$ and $\mathrm{B_{\Phi}}$ in Figure~\ref{fig:Geometry}. 
The dipole field is calculated by extracting just the $\ell=1$ component of the spherical harmonic decomposition: 
\begin{eqnarray}
\hat{M}_\ell &=&
\sqrt{\sum_{m=-\ell}^\ell \left|\int Y_{\ell m}^*(\theta,\varphi) B 
\,\mathrm{d} \Omega\right|^2}.
\end{eqnarray}

Close to the end of the simulation, the RMS field strength appears almost flat throughout the simulated domain, varying only between $\mathord\approx5\times10^{9}\mathrm{G}$ and $10^{10}\mathrm{G}$. The dipole component of the magnetic field reaches about one-third of the total field strength in the inner oxygen and neon burning shells, which are no longer convective at this point. Further out, however, the dipole is weaker by comparison to the RMS-averaged field. The slowly rotating and still convective carbon shell may be concentrated in smaller scale field structures that are similar to the non-rotating convective shell presented in \citet{VarmaMuller2021}. However, as the carbon shell has only completed 1--2 convective turnovers, it is difficult to say if this structure will be maintained at later times. 

We attempt to better visualise the evolution of the magnetic field geometry in Figure~\ref{fig:Geometry} by plotting the zonal averages of the three field components, $\mathrm{B_r, B_{\Theta}}$ and $\mathrm{B_{\Phi}}$ at two times, up to $\mathrm{15,000\,km}$. The left halves of each plot show the averages at $\approx250\mathrm{s}$, soon after magnetic field saturation is first achieved and convection is suppressed in the Neon and Oxygen shells. The right halves are at the same time chosen for Figure~\ref{fig:Dipole}, at $\approx480\mathrm{s}$, near the end of the simulation. We find that during the convective dynamo phase, the field strength is concentrated in small-scale structures, but once convective and rotational velocities are reduced, the small-scale magnetic field structures appear to diffuse away, leaving a mostly dipolar magnetic field. Some smaller-scale structures are still present, however, it isn't clear if these will still be present at even later times.

The carbon shell behaves somewhat differently from its neighbours as the shell is much larger, and already more slowly rotating at the beginning of the simulation (Figure~\ref{fig:Initial_Entropy}). Due to the slower rotation and density of the shell, the magnetic fields in the carbon shell saturate at a lower field strength. But unlike the two inner shells, the magnetic stresses here always remain below the radial kinetic stresses (Figure \ref{fig:Stress}), so convection continues unimpeded by the magnetic fields, and coupled with differential rotation, sustains a relatively constant magnetic field strength. Unfortunately, due to the very long convective turnover times, we are only able to resolve about one convective turnover in this shell. Pushing this simulation further becomes untenable as the convection in the carbon shell has begun to interact strongly with our outer domain boundary.

\subsection{Impact of magnetic fields on convection and energy generation}\label{subsec:Convection&Energy}

As we already briefly mentioned above, the amplification of the magnetic fields in our simulation leads to a very rapid suppression of convection, as well as fast transport of angular momentum out of the affected shells. Here, we attempt to understand the consequences of these dynamical changes by comparing the MHD simulation to a purely hydrodynamical simulation of this progenitor. 

As the magnetic field grows, it eventually becomes strong enough to affect the bulk flow in the convection zones. To illustrate this, we compare the spherically-averaged diagonal components of the kinetic (Reynolds) and magnetic (Maxwell) stress tensors $R_{ij}$ and $M_{ij}$. $R_{ij}$ and $M_{ij}$ are computed as
\begin{eqnarray}
R_{ij}&=& \langle \rho v_i v_j\rangle, 
\label{eqn:Stress1}\\
M_{ij}&=& \frac{1}{8\pi}\langle B_i B_j\rangle,
\label{eqn:Stress2}
\end{eqnarray}
where angled brackets denote volume-weighted averages. Note that we do \emph{not} subtract the mean rotational flow for $R_{\phi\phi}$ here.
In Figure~\ref{fig:Stress}
we present the stresses in the MHD model at $\mathord{\approx} 180\,\mathrm{s}$, where the Maxwell stresses begin to be comparable to the radial Reynolds stress, and near the final time-step of the simulation at $\mathord{\approx} 480\,\mathrm{s}$. 

\begin{figure*}
\centering
\begin{subfigure}[a]{0.95\textwidth}
   \includegraphics[width=1\linewidth]{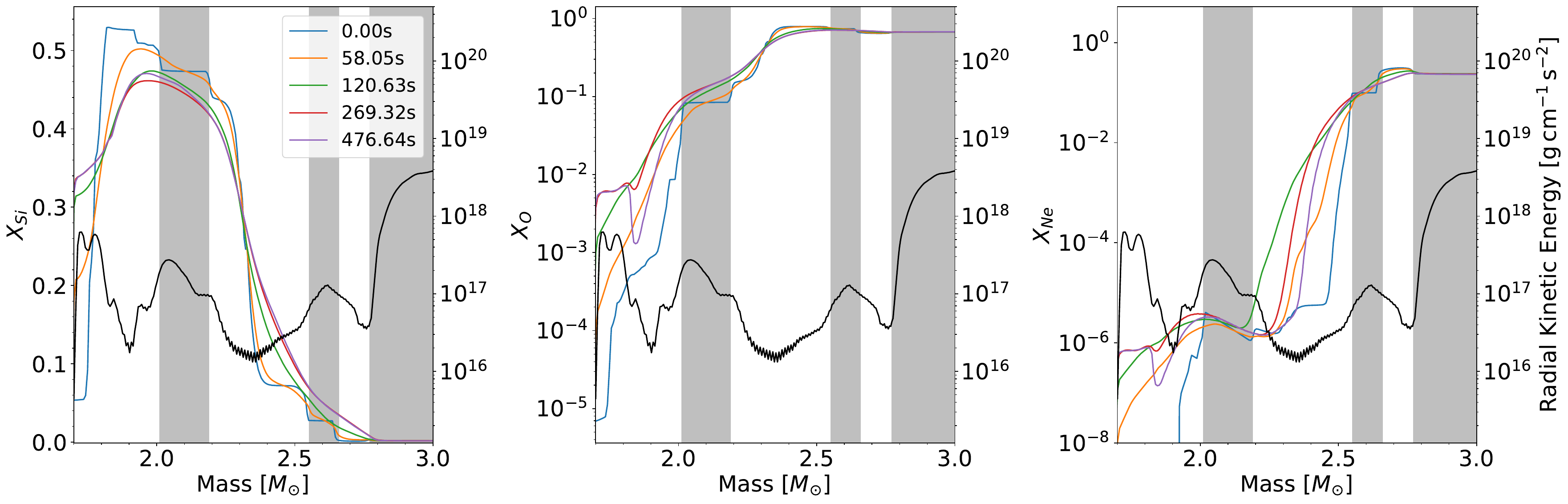}
%   \caption{}
   \label{fig:MassFrac_mag} 
\end{subfigure}

\begin{subfigure}[b]{0.95\textwidth}
   \includegraphics[width=1\linewidth]{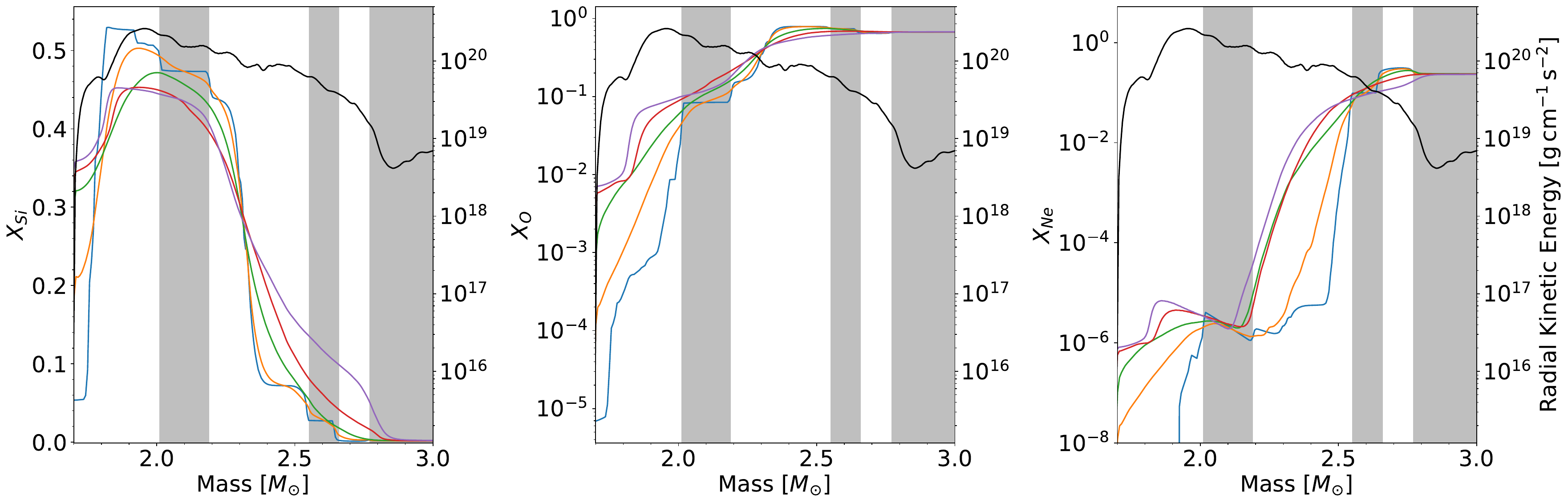}
%   \caption{}
   \label{fig:MassFrac_hydro}
\end{subfigure}
\caption{\normalsize Profiles of silicon, oxygen and neon mass fractions as a function of enclosed mass at a range of time snapshots. The top row is for the MHD case, and the bottom row for the purely hydrodynamic model. The RMS averaged radial kinetic energy at $\mathord\approx476\mathrm{s}$ is plotted in black for both MHD and hydrodynamic models. The grey bands represent the locations of the convective oxygen, neon and carbon shells when the simulation begins}
\label{fig:MassFrac} 
\end{figure*}

In Figure~\ref{fig:Stress}(a), the radial and meridional magnetic stresses are comparable to the radial kinetic stresses in the innermost regions of the star. This corresponds to pseudo-equipartition of these stress components throughout the oxygen and neon burning shells out to a mass coordinate of $\mathord{\approx} 2.8\ \msun$. The magnetic stresses then generate backreaction against the convective flows in these shells. As shown in Figure~\ref{fig:Stress}(b) at a later time in the simulation, the backreaction greatly suppresses the convective velocities in these shells, lowering the radial kinetic stresses by several orders of magnitude. We plot the RMS angle-averaged radial kinetic energy near the end of the simulation ($\mathord\approx480\,\mathrm{s}$) in Figure~\ref{fig:MassFrac} for both the MHD model (top row) and for the purely hydrodynamic case (bottom row). At this time, the suppression of convective motions in the oxygen and neon shells by strong magnetic fields cause the radial kinetic energy in the inner $2.8 \msun$ to be about three orders of magnitude lower than what is seen in the hydrodynamic model ($\mathord\approx 10^{16}-10^{17}\mathrm{g\,cm^{-1}\,s^{-2}}$ compared to $\mathord\approx 10^{19}-10^{20}\mathrm{g\,cm^{-1}\,s^{-2}}$). As mentioned above, the carbon shell has only had time for about one convective turnover, i.e., convection is yet to reach a fully developed state. At the end of our simulation, the carbon shell is still convective, and the kinetic stresses in this shell remain higher than the magnetic stresses.

At early times, we see that the angular Reynolds stresses are the dominant components due to the rapid rotation. This along with the sharp gradients in these stresses that develop means that the shear instabilities can efficiently mix material more efficiently than in the underlying 1D stellar evolution models, where there is little mixing beyond the convective zones on dynamical time scales.
Shear mixing outside the convective regions initially plays a significant role in the MHD model as well. The $R_{rr}$ stress component, which is indicative of radial motions that contribute to turbulent mixing, stays high outside the convective regions in the MHD model initially
(Figure~\ref{fig:Stress}(a)).
However, in the MHD model, the transport of angular momentum flattens the rotation profile
(as discussed in detail in Section~\ref{subsec:Rotation})
, which we can see from the change in $R_{\theta\theta}$ and $R_{\phi\phi}$, significantly reducing the shear mixing at late times. In our purely hydrodynamic model, however, the Reynolds stresses remain roughly the same throughout our simulation, leading to continuous enhanced mixing compared to the MHD counterpart. We find that this enhanced shear mixing compared to the initial expectation from the 1D stellar evolution model means that the burning occurs at different regions, outside the initial location of the convection zones.

The consequence of this can be seen by analysing how the mass fractions of several key elements evolve. In Figure~\ref{fig:MassFrac}, we compare the mass fractions of silicon, oxygen and neon in the MHD simulation and the purely hydrodynamic simulation. We also plot the radial kinetic energy at the end of the simulations, to further stress the differences in turbulent mixing when magnetic fields are introduced. The plots are limited to the inner $3 \msun$ of the enclosed mass to focus on the oxygen and neon burning shells, which are most strongly affected by magnetic fields.   

The radial profiles of all three elemental mass fractions in the hydrodynamic and MHD model evolve very similarly at early times (up to $\mathord\approx120\,\mathrm{s}$ in Figure \ref{fig:MassFrac}), when the magnetic fields are not strong enough to significantly affect mixing. At later times, however, the differences in mixing become quite apparent. The plots of the  silicon mass fraction show that large fractions of silicon are mixed outwards to an enclosed mass of $\mathord\approx2.65\msun$ in the hydrodynamic model, while there is little mixing beyond $\mathord\approx2.5\msun$ in the MHD simulation.

The inhibition of mixing also means that material that gets burnt at the base of a shell is less efficiently replenished by fresh material or not replenished at all.
This effect is seen in the oxygen shell as it burns material in a relatively narrow region around $1.85 \msun$. In the MHD simulation, a sharp drop in the oxygen and neon mass fractions develops in this region, which gets steeper at later times due to the rapid burning (mostly of oxygen), which is no longer replenished by convective and shear mixing. It is clear that this is caused by the very strong magnetic fields and a significant reduction in turbulent mixing by comparing the mass fractions to the purely hydrodynamic simulation. Without magnetic fields, there is no sharp drop in mass fraction at the bottom of the burning shell. The profiles remain smoother within the shell and are even smoothed beyond the boundaries by shear mixing. The oxygen and neon mass fractions in the oxygen burning region even increase
over time, as the rapid rotation and convection act to continually entrain fresh material into the oxygen and neon shells. 

\begin{figure*}
\centering
    \subfloat[]{\includegraphics[width=0.45\linewidth]{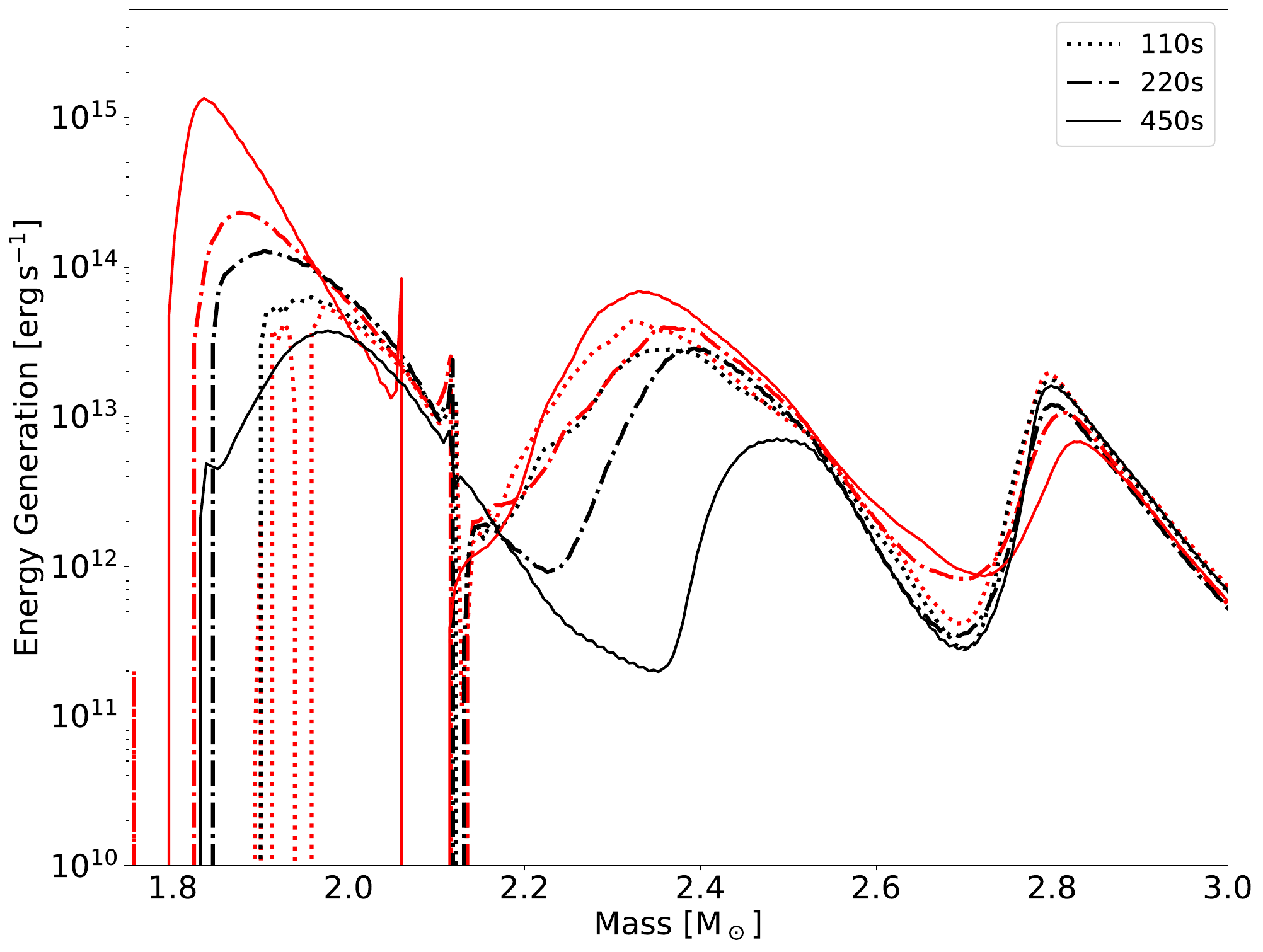}}
\hfil
    \subfloat[]{\includegraphics[width=0.45\linewidth]{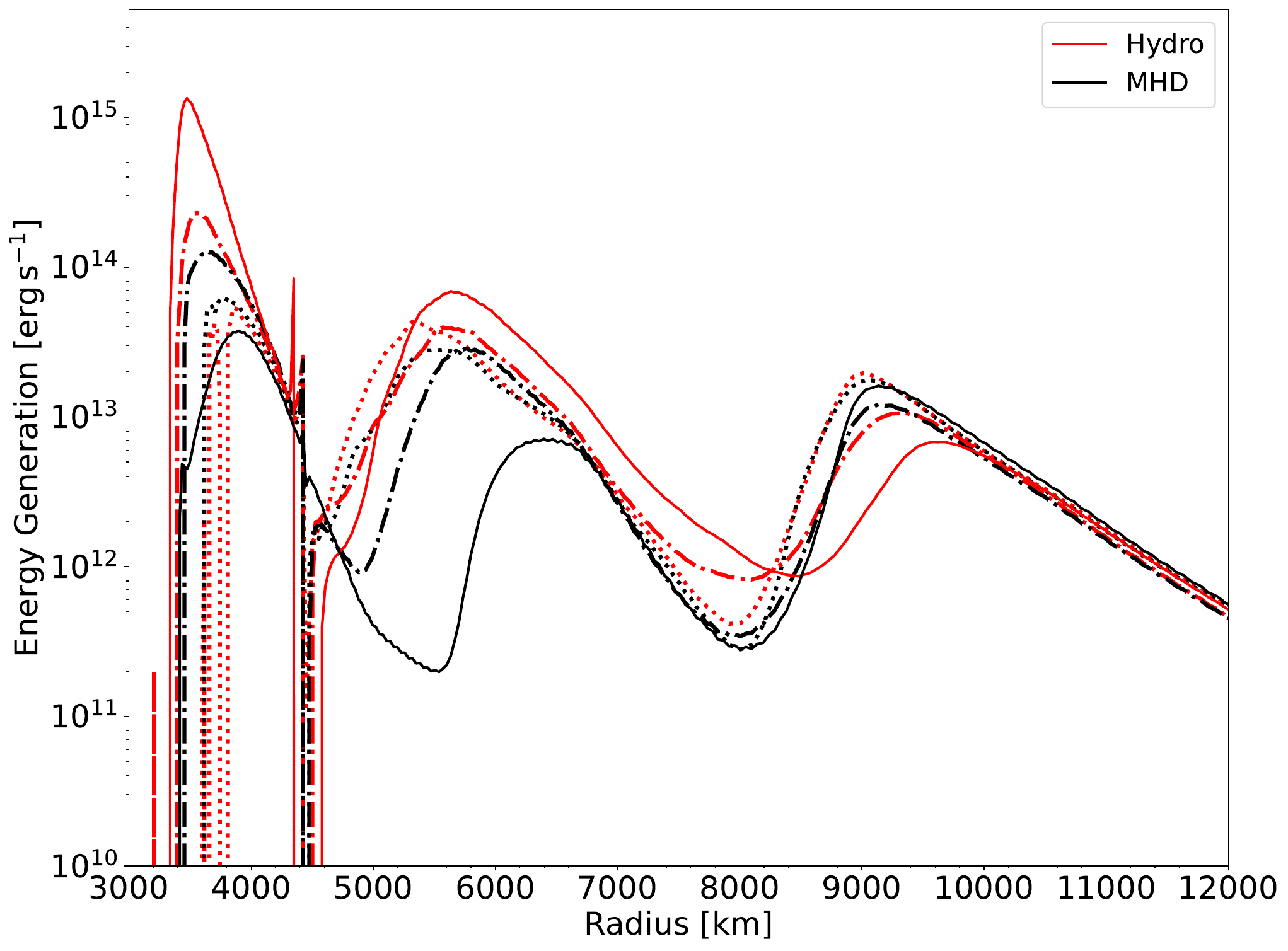}}

\caption{\normalsize Profiles of the nuclear energy generation rate at $110\, \mathrm{s}$ (dotted), $220\, \mathrm{s}$ (dashed) and $450\, \mathrm{s}$ (solid) for both the MHD (black) and hydrodynamic (red) simulations. The energy generation rates are presented both as a function of mass in panel (a) and as a function of radius in panel (b). }
    \label{fig:EnergyGeneration_profile}
\end{figure*}

We also find reduced mixing in the neon burning shell between $\mathord\approx2.20\msun$ and $2.40\msun$. Here, instead of a sharp drop, we see the steep gradient of neon, where neon is consumed, move outwards after the initial transient where material is mixed inwards.
This has the effect of shifting the location of the base of the burning shell. We show this phenomenon more clearly in Figure~\ref{fig:EnergyGeneration_profile}, which presents angle-averaged radial profiles of the energy generation rate at $110\, \mathrm{s}$, $220\,\mathrm{s}$ and $450\,\mathrm{s}$ (dotted, dashed and solid lines, respectively). We show their evolution in both mass and radial coordinates in Figures~\ref{fig:EnergyGeneration_profile}(a) and (b) respectively, from the inner boundary of our simulation $1.75\msun$ to an enclosed mass of $3.0\msun$, and radius of $3000\,\mathrm{km}$ to $12000\,\mathrm{km}$ for both the hydrodynamic model and the MHD model. The energy generation profiles for both models show three clear peaks, which correspond to the three burning shells (oxygen, neon and carbon). We see that both simulations quickly deviate from each other, with the energy generation rate in the oxygen and neon shells in the hydrodynamic model receding backwards, and getting stronger at later times, while the opposite is seen in the MHD simulation. We note that the relative change in the position of these energy generation peaks are largely similar in both mass and radial coordinates.

\begin{figure}
 \includegraphics[width=\columnwidth]{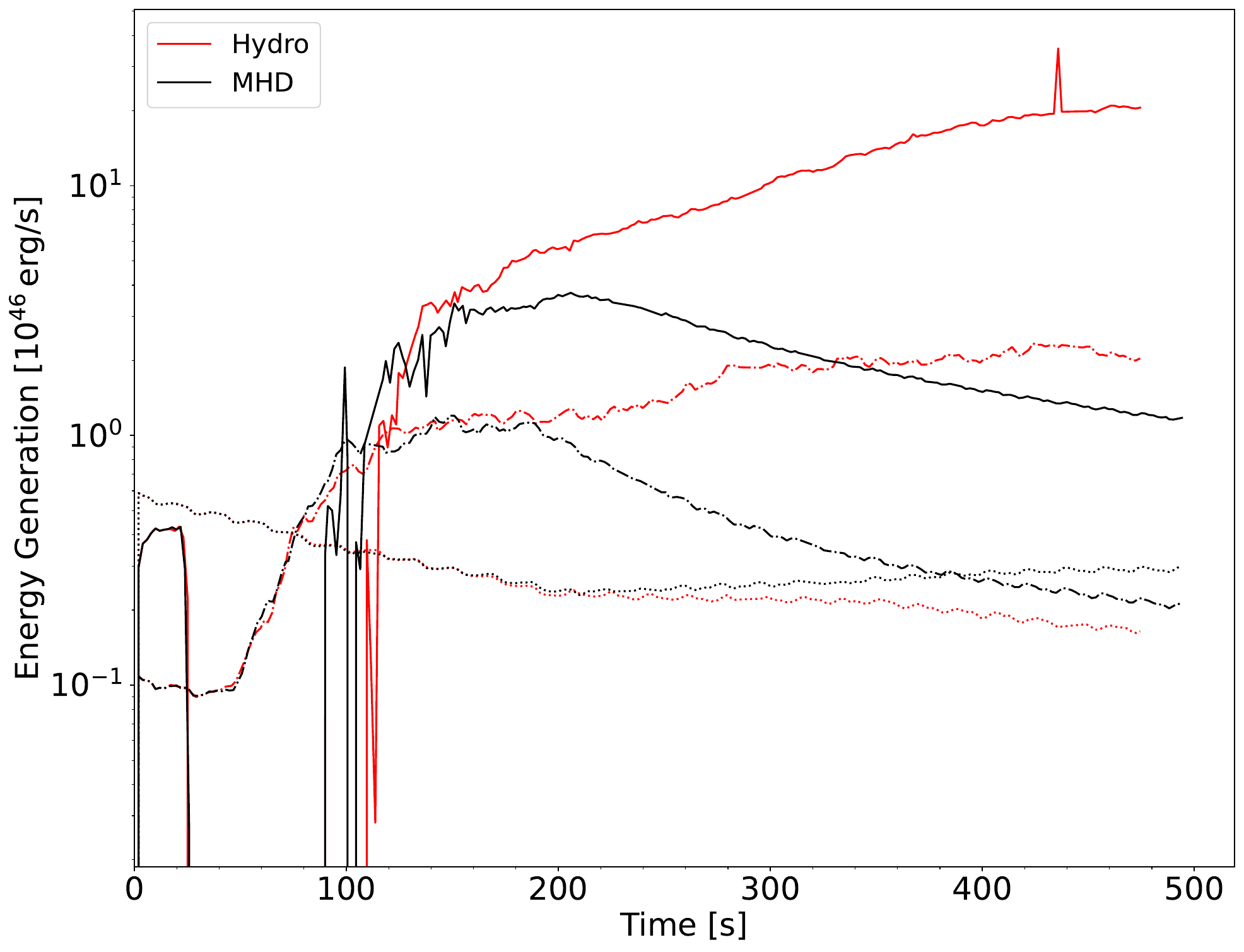}
 \caption{\normalsize Volume-integrated nuclear energy generation rate for the oxygen (solid), neon (dashed) and carbon (dotted) burning shells with time for both the MHD (black) and hydrodynamic (red) simulations.}
 \label{fig:EnergyGeneration}
\end{figure}

Figure~\ref{fig:EnergyGeneration_profile} clearly shows that the second energy generation peak in the MHD model, which corresponds to neon shell burning, moves from $\mathord\approx 2.35\msun$ ($\mathord\approx 5400\,\mathrm{km}$) at early times to $\mathord\approx 2.50\msun$ ($\mathord\approx 5800\,\mathrm{km}$) later on, compatible with the change in the neon mass fraction profiles in Figure~\ref{fig:MassFrac}. Due to the lack of turbulent mixing, the peak of neon burning moves radially outward as neon gets burnt at its initial position. However, since the neon burning rate is extremely temperature-sensitive ($\propto T^{50}$, \citealt{Woosley2002}), the energy generation rate drops significantly when burning is moved to a slightly cooler region.

We see a similar effect for the oxygen shell. At early times, before the magnetic field heavily suppresses turbulent mixing, both the MHD and hydrodynamic model mix material rapidly. The convective boundary mixing (CBM), particularly at the lower boundary of the oxygen shell, entrains material, increasing the size of the oxygen shell, causing it to move inwards (both in mass and in radius). We see the peak energy generation move from $\mathord\approx 1.95\msun$ ($\mathord\approx 3900\,\mathrm{km}$) to $\mathord\approx 1.85\msun$ ($\mathord\approx 3500\,\mathrm{km}$).

After turbulent convection is suppressed in the MHD simulation, however, the nuclear energy generation peak in the oxygen shell behaves similarly to the neon burning shell, i.e., it  moves outward in mass and radius to where oxygen is burned at a lower temperature. The hydrodynamic model, however, continues to entrain material from beneath the oxygen shell, moving the peak of its energy generation deeper into the core.  oxygen burning is also a very temperature-sensitive reaction ($\propto T^{33}$,  \citealt{Woosley2002}), so these shifts in the burning shells lead to noticeable changes in the total energy generation rate over time.

The consequences of the suppression of mixing in the MHD model are seen more clearly in Figure~\ref{fig:EnergyGeneration}. Here, we plot the total (volume-integrated) energy generation in the oxygen, neon and carbon burning shells with time for both the hydrodynamic and MHD models. As we see in Figure~\ref{fig:EnergyGeneration_profile}, the lack of mixing in the MHD model leads to the location of the oxygen and neon shells moving radially outwards with time, to cooler regions of the star. This causes a gradual decrease in energy generation after the magnetic fields in these shells first achieve saturation at $\mathord\approx200\,\mathrm{s}$. The increased mixing in the hydrodynamic model, on the other hand, causes a subsequent increase in peak energy generation rate as the shells move deeper into the star.

Interestingly, after $\mathord\approx 220\,\mathrm{s}$ of similar energy generation in the carbon shell in both models, the energy generation rate starts to increase in the MHD model and decreases in the hydrodynamic case. This is likely due to the slight change in the position of the carbon burning shell for the hydrodynamic model while it remains mostly stationary in radius in the MHD model. From the profiles in Figure ~\ref{fig:EnergyGeneration_profile}, this is likely caused by the radial expansion of the neon shell in the hydrodynamic model as its energy generation rate increases, pushing the carbon shell further outwards. 
In the MHD model, by contrast, the energy generation rate in the neon shell 
has dropped over time and hence there the carbon shell is not driven outward.

% -------------------------------------------------------------------------------------------
\subsection{Evolution of rotation and angular momentum transport}
\label{subsec:Rotation}

In addition to its effect on turbulent mixing, the development of strong magnetic fields leads to rapid redistribution of angular momentum. 
The evolution equation for the angular momentum can be obtained by taking the cross product of the position vector $\mathbf{r}$ with the fluid momentum equation \citep{Shu1992, Mestel1999}.
When including magnetic stresses in the momentum equation and integrating over spherical shells, one obtains \citep[cf.\ ][]{Charbonneau1993, Spruit2002},

\begin{equation}
\frac{{\partial}
\langle\rho v_\phi r \sin \theta \rangle}{{\partial}t} - 
{\nabla_r}\cdot
\left\langle\rho v_r v_\phi r \sin \theta  + 
\frac{{B_r}{B_\phi}}{4\pi}r \sin \theta \right\rangle = 0,
\end{equation}
\label{eq:AngMom}
where $\nabla_r$ is the radial component of the divergence operator and angled brackets denote averages over solid angle.
We then perform a Reynolds/Favre decomposition \citep{Favre1965} around a base state with constant angular velocity $\Tilde{\Omega}_z$, 
\begin{equation*}
    \Tilde{\Omega}_z = \frac{\langle \rho \Omega_z r^2 \sin^2 \theta \rangle}{\langle \rho r^2 \sin^2  \rangle} = \frac{\Tilde{v}_\phi r \sin \theta}{\Tilde{i}_{zz}},
\end{equation*}
on spheres, as in \citet{McNeill2022}. Here $\Tilde{i}_{zz} = \langle \rho r^2 \sin^2  \rangle/\Hat{\rho}$. Note that we use hats and primes for volume-weighted Reynolds averages and their fluctuating components:

\begin{equation*}
    \Hat{X}(r) = \langle X \rangle = \frac{1}{4\pi}\int X \,\mathrm{d}\omega \\
    X'(r,\theta,\phi) = X - \Hat{X},
\end{equation*}
where  $d\omega = \sin \theta \,\mathrm{d}\theta \,\mathrm{d}\phi$ is the solid angle element. We denote mass-weighted averages and fluctuations with tildes or angled brackets and double primes:

\begin{equation*}
    \Tilde{X}(r) =  \frac{\int \rho X d\omega}{\int \rho d\omega} \\
    X''(r,\theta,\phi) = X - \Tilde{X}.
\end{equation*}

Applying the usual rules for Favre averages,
$\langle\rho X\rangle = \hat{\rho}\Tilde{X}$, $\langle\rho \Tilde{X}\Tilde{Y}\rangle = \hat{\rho}\Tilde{X}\Tilde{Y}$ and $\langle\rho \Tilde{X}Y''\rangle = 0$, we get
\begin{equation}
\begin{split}
\frac{\partial \Hat{\rho}\Tilde{\Omega}_z\Tilde{i}_{zz}}{\partial t} &+ 
\nabla_r \cdot (\langle\rho \Tilde{v}_r(\Tilde{\Omega}_z + \Omega''_z) r^2 \sin^2 \theta\rangle \\
& +\, \langle \rho v''_r\Tilde{\Omega}_z r^2 \sin^2 \theta \rangle
+\, \langle \rho v''_r \Omega''_z r^2 \sin^2 \theta \rangle \\
% & +\, \langle\frac{\Tilde{B}_r\Tilde{B}_\phi}{4\pi}r \sin \theta \rangle 
&-\, \langle\frac{B_r B_\phi}{4\pi}r \sin \theta \rangle ) = 0.
\end{split}
\label{eq:Favre_AngMom}
\end{equation} 

We see from the decomposed angular momentum Equation~(\ref{eq:Favre_AngMom}) that angular momentum is transported by four distinct flux terms. In the order they are listed above we have an advective term, a meridional circulation term and turbulent transport, as well as an additional magnetic stress term. From this equation, aside from the usual hydrodynamic terms, the radial flux of angular momentum also depends on the strength of the magnetic fields. 

\begin{figure*}
    \centering
    \begin{subfigure}[b]{0.475\textwidth}
        \centering
        \includegraphics[width=\textwidth]{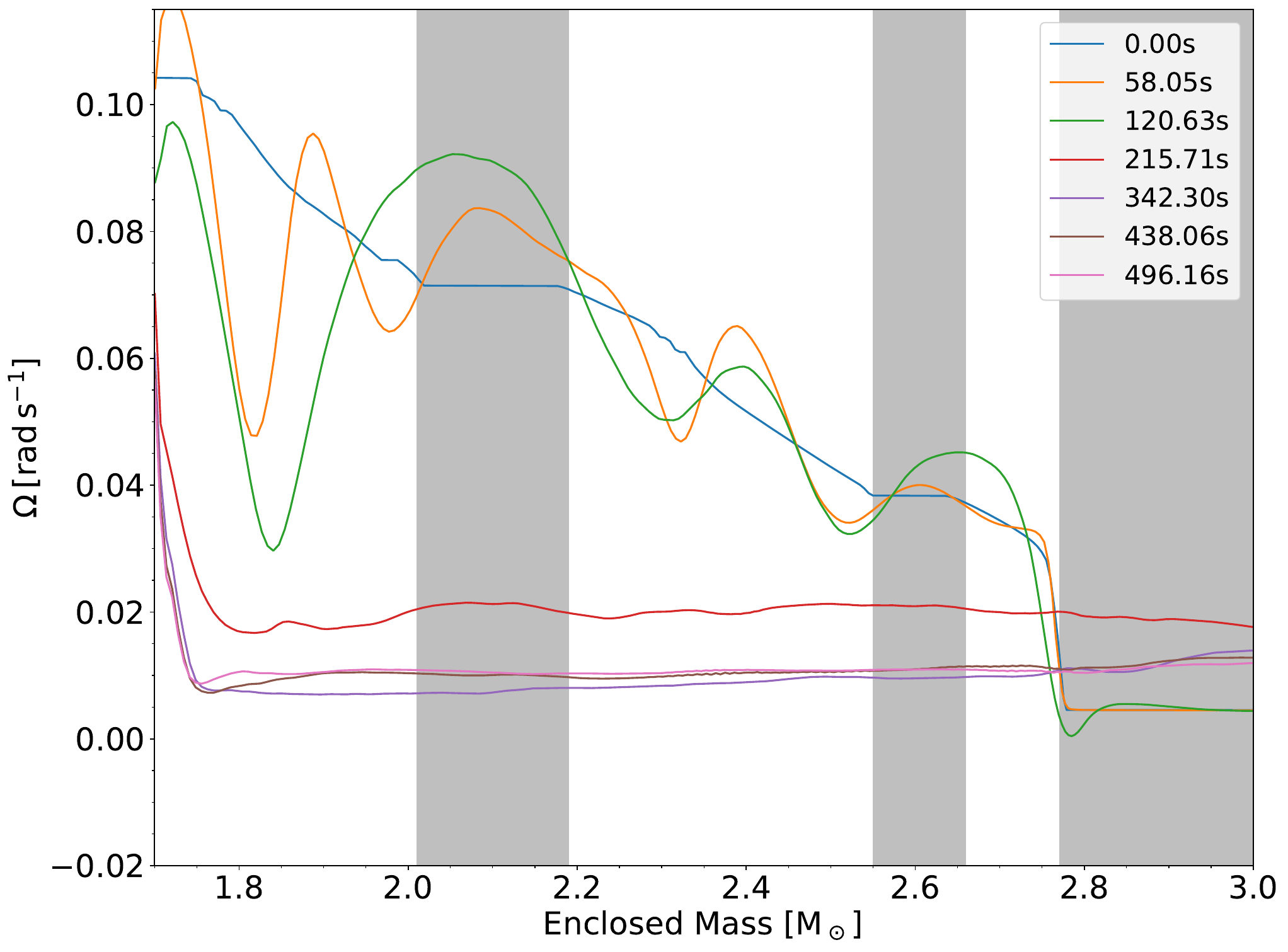}
        \caption[Network2]%
        {{\small MHD Angular Velocity}}    
        \label{fig:mean and std of net14}
    \end{subfigure}
    \hfill
    \begin{subfigure}[b]{0.475\textwidth}  
        \centering 
        \includegraphics[width=\textwidth]{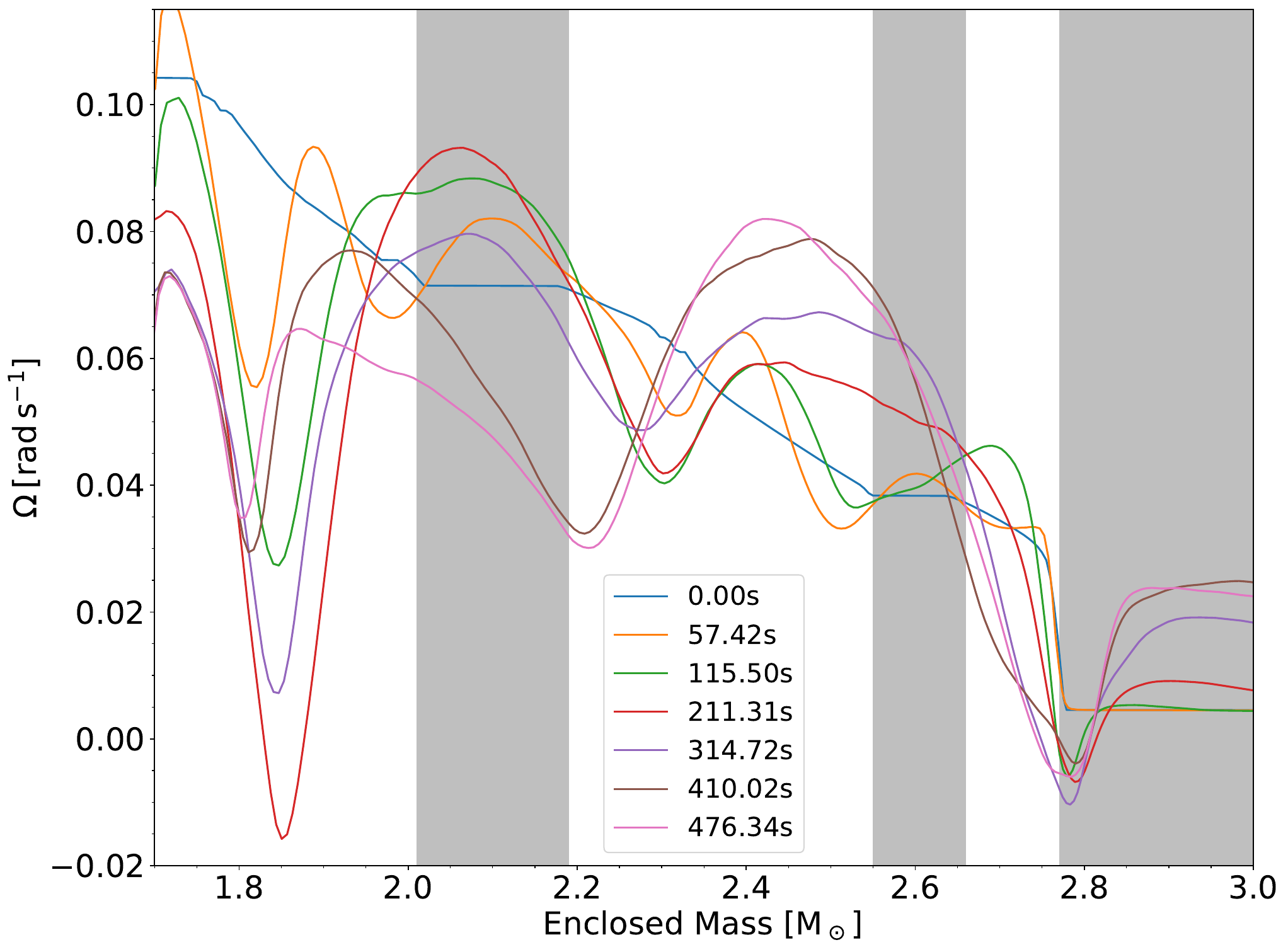}
        \caption[]%
        {{\small Hydro Angular Velocity}}    
        \label{fig:mean and std of net24}
    \end{subfigure}
    \vskip\baselineskip
    \begin{subfigure}[b]{0.475\textwidth}   
        \centering 
        \includegraphics[width=\textwidth]{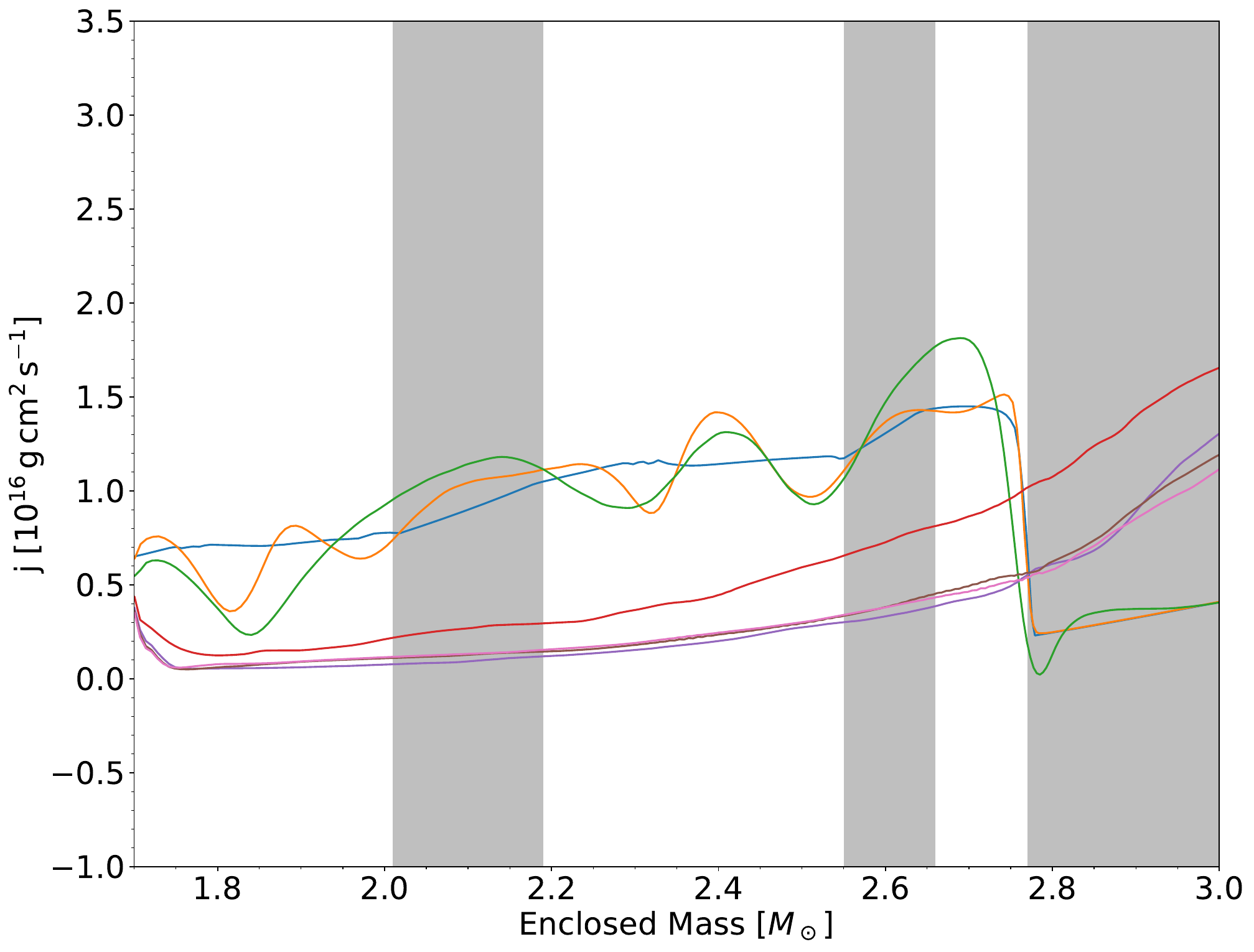}
        \caption[]%
        {{\small MHD Angular Momentum}}    
        \label{fig:mean and std of net34}
    \end{subfigure}
    \hfill
    \begin{subfigure}[b]{0.475\textwidth}   
        \centering 
        \includegraphics[width=\textwidth]{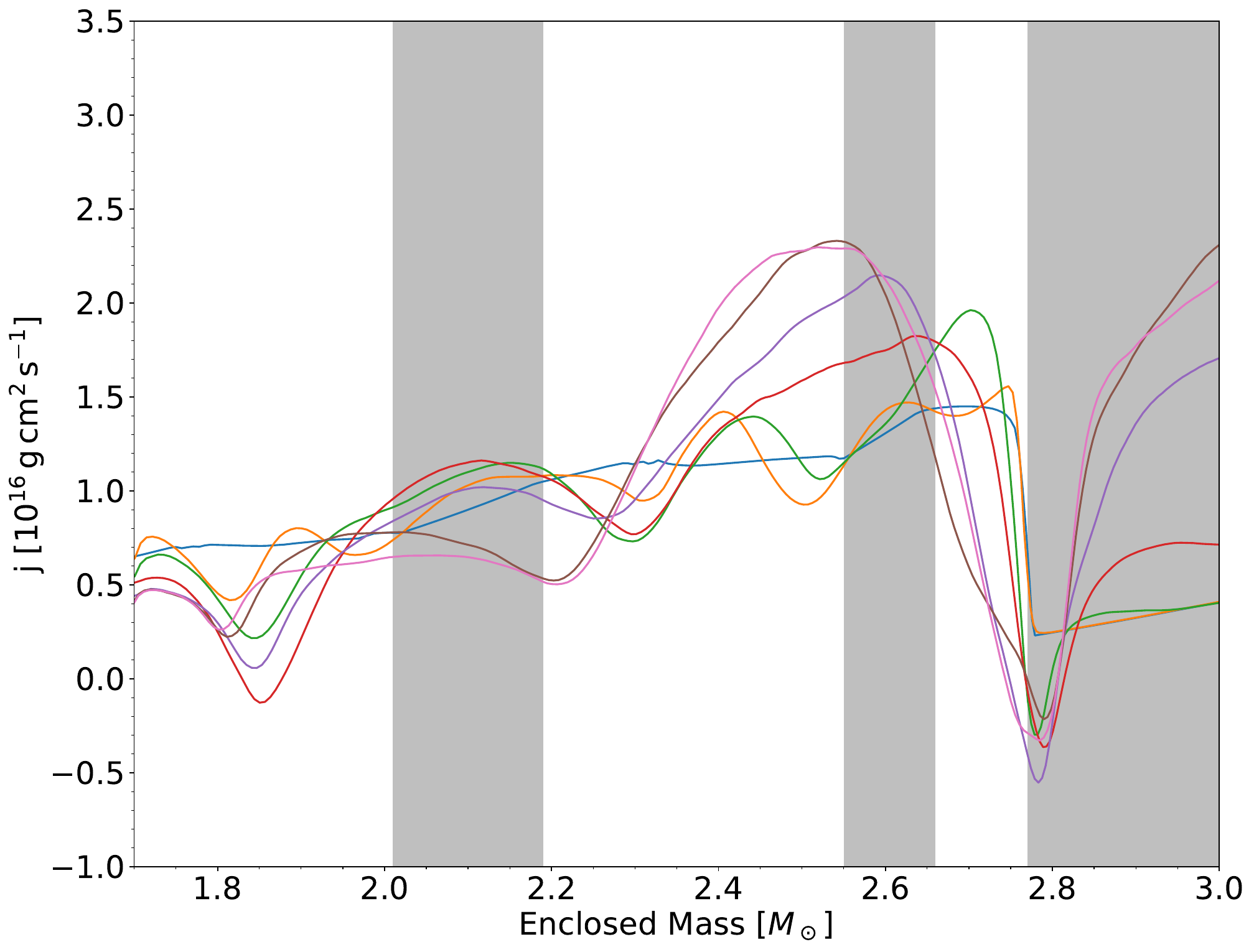}
        \caption[]%
        {{\small Hydro Angular Momentum}}    
        \label{fig:mean and std of net44}
    \end{subfigure}

    \caption{\normalsize Profiles of angular velocity (top row) and angular momentum (bottom row) as a function of enclosed mass for different times during the simulation. Plots on the left column are for the MHD case, and the right column for the purely hydrodynamic model. The plots are limited to mass coordinates inside $3\msun$ where the most clear dynamical differences occur.  The shaded bands represent the locations of the convective oxygen, neon and carbon shells when the simulation begins.}
    \label{fig:Omega}
\end{figure*}

The evolution of the angle-averaged rotation rates in the MHD model and purely hydrodynamic model are depicted in Figure~\ref{fig:Omega}. The top row shows the rotation rates, $\Tilde{\Omega}$, over time for both models, and the bottom row presents angle-averaged specific angular momenta. We note that in comparison to its hydrodynamic counterpart, as expected, the MHD model exhibit more efficient angular momentum transport. At the innermost region of our domain, the rotation rate is lowered by over an order of magnitude due to outward transport of angular momentum, from over $\mathrm{0.10 \, rad\,s^{-1}}$ initially to $\mathord{\approx} 0.01 \, \mathrm{rad}\,\mathrm{s}^{-1}$). The rotation profile is flattened considerably.
The angular momentum is taken up by carbon shell outside $2.8 \msun$, whose rotation rate increases. We note that Figure~\ref{fig:Omega} only shows a limited inner portion of the total simulated carbon shell.

The hydrodynamic model displays much less smooth rotation profiles than the MHD model. It is  noteworthy that at the various convective shell boundaries, the rotation profile shows significant dips. Figure~\ref{fig:Omega}(b) shows dips in rotation at $\mathord\approx\mathrm{1.85\msun}$, which is the base of the oxygen shell, between the oxygen and neon shells at a location that varies over time between $\mathord\approx\mathrm{2.20\msun}$ and $\mathord\approx\mathrm{2.40\msun}$, and finally between the neon and carbon shells at $\mathord\approx\mathrm{2.80\msun}$. 
It is particularly interesting that some of these dips even reach negative values, i.e., there are shells with net retrograde rotation, which remain quite stable.
On closer inspection of the rotation profile of the MHD model, we see that these dips in the rotation profile also begin to form early on in this case (Figure~\ref{fig:Omega} (a) at 120s). However, the rotation profile is quickly smoothed once angular momentum transport due to magnetic fields become efficient. 

Unlike in the MHD model, where angular momentum is simply transported outward, the redistribution of angular momentum in the hydrodynamic model appears more complicated (Figure~\ref{fig:Omega} (d)). Effectively, positive angular momentum is transported into convective shells from the shell boundaries. This leads to an interesting non-monotonic rotation profile, where the fastest 
rotation rate is not reached at the inner boundary, but instead at $2.4\texttt{-}2.5 \msun$  at late times, as well as the aforementioned dips between the convective shells. Redistribution of angular momentum also increases the rate of rotation in the inner carbon shell (around $2.8 \msun$) and 
induces strong (radial) differential rotation there.

\begin{figure*}
\centering
    \subfloat[]{\includegraphics[width=0.45\linewidth]{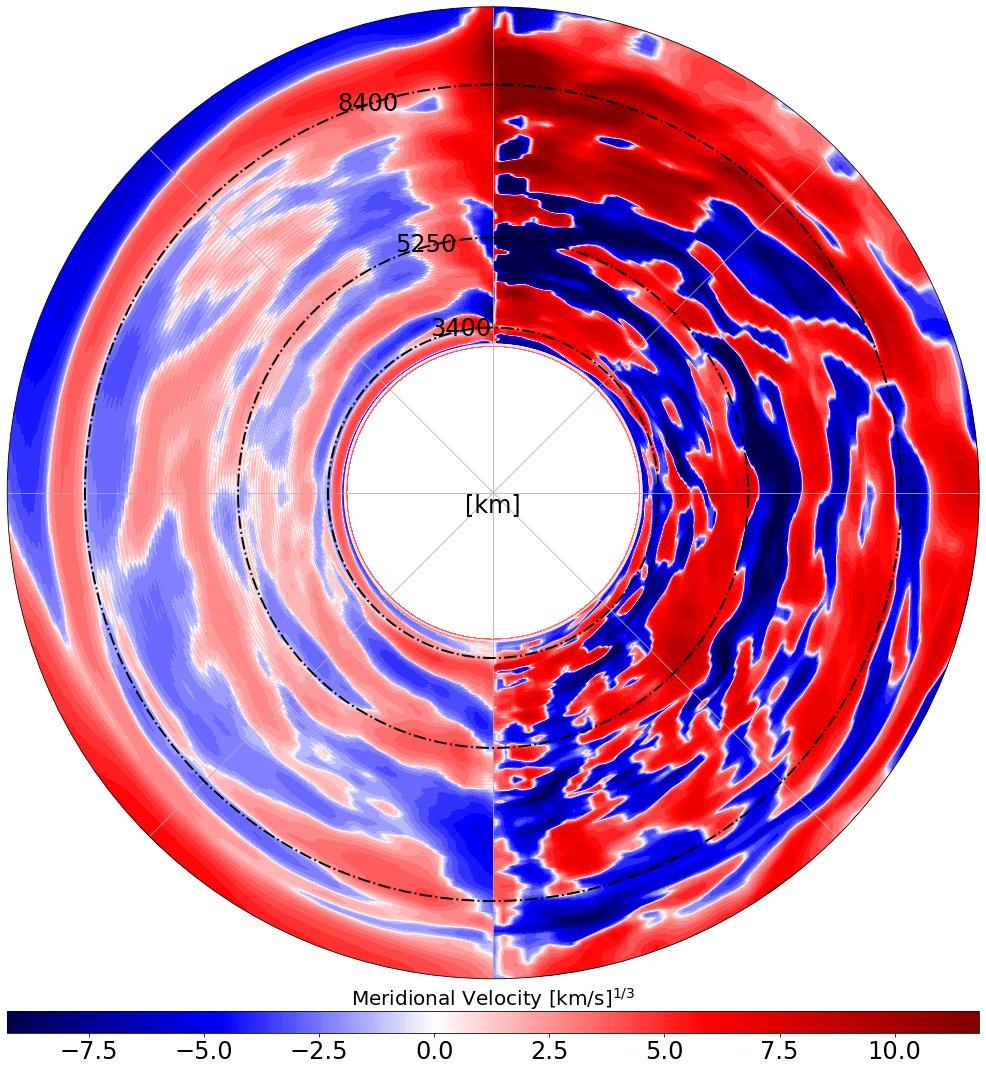}}
\hfil
    \subfloat[]{\includegraphics[width=0.45\linewidth]{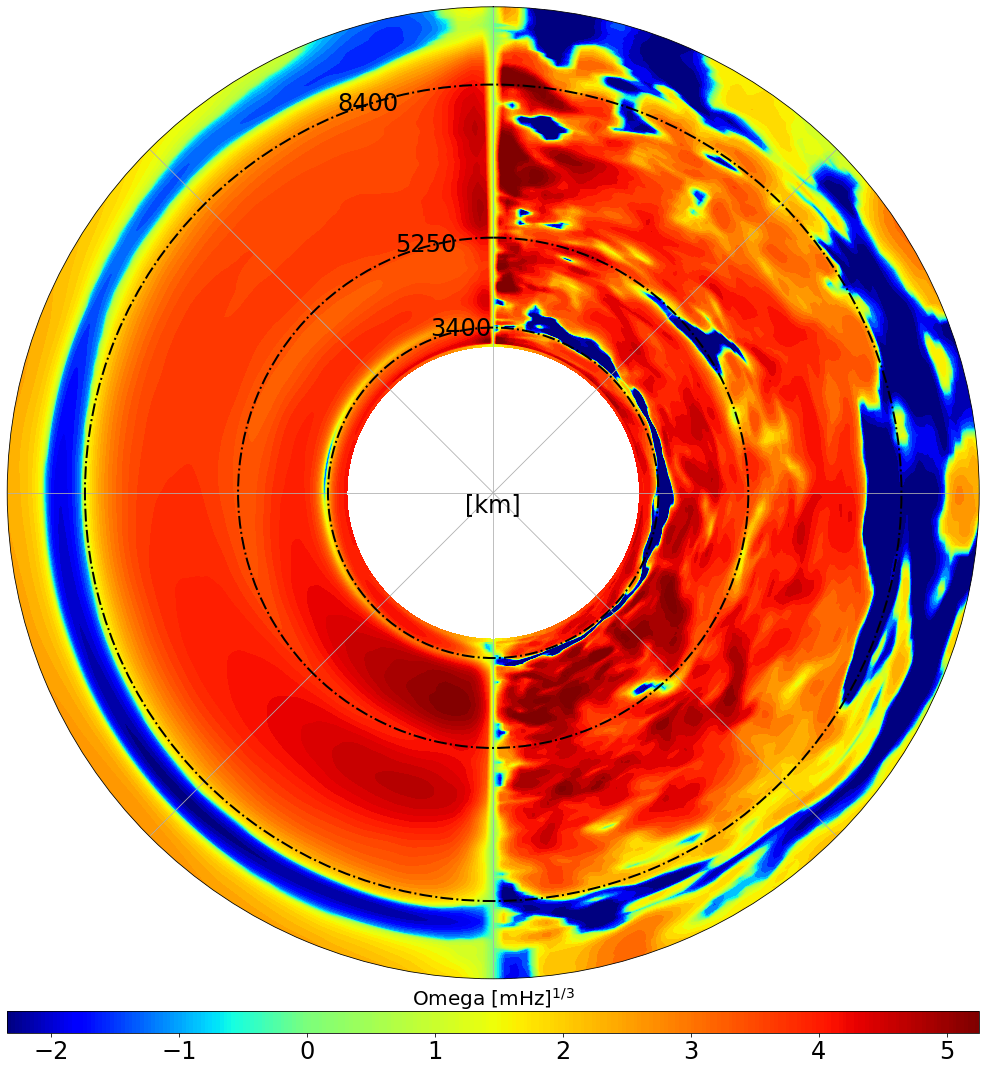}}

\caption{\normalsize The left halves of panels (a) and (b) depict zonal and time average quantities from $250\,\mathrm{s}$ to $400\,\mathrm{s}$, while the right halves are meridional snapshots at $300\,\mathrm{s}$. The cube root of meridional velocity is shown in panel (a) and the rotation rate in panel (b). The positive meridional velocities represent clockwise motion. Dashed-dotted black lines mark the approximate initial radii of the base of the oxygen, neon and carbon shells.}
    \label{fig:2D_AVG}
\end{figure*}

To better understand the emerging rotation pattern in the hydrodynamic model and the counterintuitive phenomenon of retrograde rotation, we consider zonal and temporal averages
of the meridional velocity and the rotation rate in Figure~\ref{fig:2D_AVG}. In Figure~\ref{fig:2D_AVG} (a), the meridional velocity plotted is an average over $\phi$ (zonal average) and time of $|\mathbf{v}_r + \mathbf{v}_\theta|$, while Figure~\ref{fig:2D_AVG} (b) plots the same averages of the angular velocity, $\Omega = v_\phi/(r \sin \theta)$, where $r$ is the radius. Both figures plot the cube root of their original values to retain the direction of the flow, but reduce the dynamic range. The positive (red) meridional velocity represents clockwise motion, with the negative (blue) flows showing counter-clockwise flows (viewed from the North pole).

The left halves of Figure~\ref{fig:2D_AVG} (a) and (b) show the zonal flow averages of the quantities, while the right halves show snapshots of the two quantities on a meridional slice. From Figure~\ref{fig:2D_AVG}(b), it is clear that the retrograde rotation occurs mostly at the base of the carbon shell, and to a lesser extent, at the base of the oxygen shell. The snapshot on the right half of Figure~\ref{fig:2D_AVG}(b) shows that regions of retrograde rotation form at the base of the neon shell as well, but these are more transient and do not show up in the zonal averages. We also observe that for the hydrodynamic model, although the retrograde rotation appears to form as a shell at the base of the carbon shell, in general, the rotation pattern that forms is \emph{not} shellular. At the base of both the oxygen and carbon shell, we rather see indications of anti-solar differential rotation with faster rotation near the poles.

Such a rotation pattern has been observed before in simulations of surface convection zones.
\citet{Featherstone2015} attribute the development of different rotation profiles, in part, to a link between the differential rotation and the meridional circulation in the convection zone. Anti-solar rotation profiles, are attributed to inward angular momentum transport, which establishes a single-celled meridional circulation profile throughout the convection zone. This circulation transports angular momentum polewards, spinning up the poles relative to the equator. 
Although our model exhibits antisolar-like rotation profiles at the base of the carbon ($8400\,\mathrm{km}$) and oxygen ($\mathord{\approx}3400\,\mathrm{km}$) shells in Figure ~\ref{fig:2D_AVG}(b), the corresponding meridional circulation in Figure ~\ref{fig:2D_AVG}(a) does not exhibit the single-celled structure expected from \citet{Featherstone2015}. We instead find that the meridional flows are not clearly structured, but appear more similar to a case with multi-celled circulation.

We see an analogous meridional circulation flow develop in the inner regions of the carbon shell of our model (above $8400\mathrm{km}$), with two large cells of material (Figure ~\ref{fig:2D_AVG}(a)). One noticeable difference shown in our model is that the circulation velocities do not drop off towards the poles, instead, they remain very strong. 
This may be part of the reason why the model exhibits a stable retrograde shell of material at $8400\,\mathrm{km}$ that extends across almost all latitudes as shown in Figure~\ref{fig:2D_AVG}(b),
rather than being confined to the equatorial region as in the surface convection models presented in \citet{Featherstone2015}.

One major difference that our simulations have that likely alters the meridional flows, is the proximity of other convective shells. While in surface convection simulations, the shell is bounded only by a radiative core, in our model, we have three adjacent convective shells. Although these shells are initially separated by thin radiative zones in the 1D stellar structure input model, the rapid rotation and turbulent convection increases the amount of mixing that occurs at the convective shell boundaries in the 3D model, causing the convective shells to start interacting. This adds additional complications to the transport of angular momentum, and in turn, to the meridional circulation.

\begin{figure}
    \centering
    \includegraphics[width=\columnwidth]{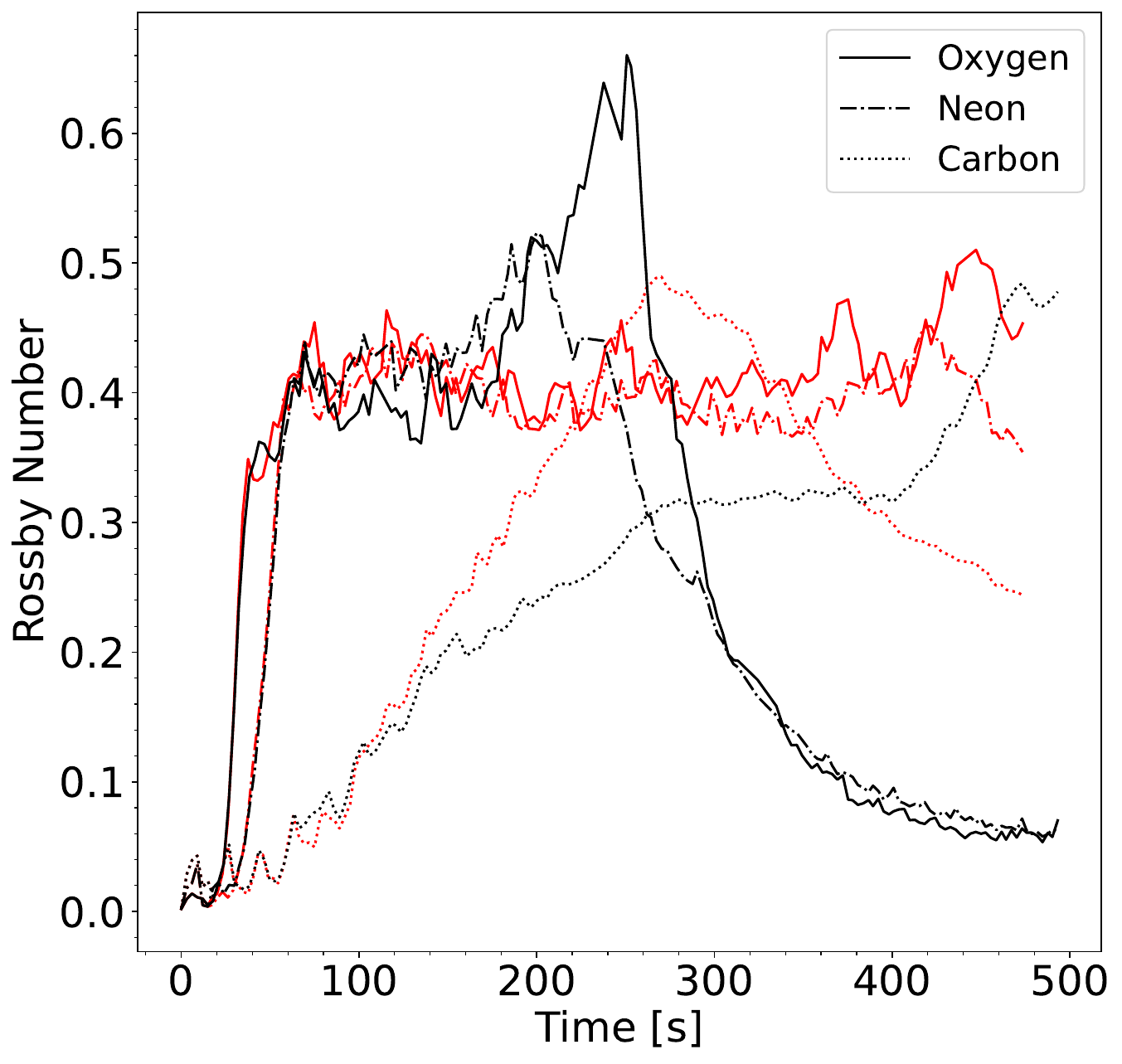}
    \caption{\normalsize Time evolution of the Rossby number in the three convective shells depicted by the different line styles for
    the oxygen, neon, and carbon shell
    both for the MHD (black) and hydrodynamic (red) simulations.}
    \label{fig:Rossby}
\end{figure}

From extensive studies in solar physics, it is often expected that the stable differential rotation in our Sun is maintained by rotating turbulent convection  \citep[e.g.,][]{ruediger_89, Hotta2022}, due to the interplay of buoyancy and inertial forces. We suspect that, similarly, the complicated rotation profile developed in our purely hydrodynamic model occurs due to an interaction between the Coriolis and buoyancy forces. Similar retrograde rotation patterns have been found in studies of rotating solar-like stars, and depend on the Rossby number of the flow \citep[e.g.,][]{Brown2008,Guerrero2013, Brun2017, Camisassa2022}. To confirm that our model is in the relevant regime where Coriolis forces are strong enough to make such an effect plausible, we analyse how the Rossby number compared to the MHD model, which develops a flat rotation profile. We define the Rossby number of any given burning shell as

\begin{equation}
    \mathrm{Ro} = \frac{v_\mathrm{conv}}{2\Omega_\mathrm{shell}\Delta r},
\end{equation}

where $v_\mathrm{conv}$ is the convective velocity, $\Omega_\mathrm{shell}$, is the rotation rate, and $\Delta r$ is the radial extent of each burning shell. The time evolution of the Rossby number in the oxygen, neon and carbon shell is shown in Figure~\ref{fig:Rossby}. We find that the Rossby number for the oxygen and neon shell for both models starts at $\mathord\approx 0.4$ (and remains largely unchanged for the hydrodynamic model), indicating that the flows are strongly shaped by the Coriolis force. The Rossby numbers in the MHD model clearly reflects that the magnetic fields starts to strongly alter the bulk flows. At $\mathord\approx 180\mathrm{s}$ the magnetic fields become strong enough to start rapidly transporting angular momentum outwards, slowing the rotation of the star and hence increasing the Rossby number. The Rossby number continues to increase until $\mathord{\approx} 200\,\mathrm{s}$ in the neon shell and $\mathord{\approx} 240\,\mathrm{s}$ in the oxygen shell,  when the suppression of convective flows by magnetic stresses becomes dominant, lowering the Rossby number.

The carbon shells of both models clearly have not reached a convective steady state and hence not even a  transient steady state in the Rossby number can be discerned. The MHD case initially transports angular momentum out of the star more efficiently. We see in Figure \ref{fig:Omega}(c) that angular momentum from the rapidly rotating inner regions (inner $2.8\msun$), is transported out to the carbon shell ($2.8 - 3.0\msun$). Focusing on the lines at $215\,\mathrm{s}$ in Figure \ref{fig:Omega}(a), we see that at these times, rotation rate in the inner carbon shell increases in the MHD case compared to the purely hydrodynamic model. This leads to the deviation in Rossby number initially seen from $\mathord\approx 150\,\mathrm{s}$, causing the Rossby number in the MHD model to be lowered compared to the hydrodynamic model. After $\mathord\approx260\,\mathrm{s}$, the Rossby number in the hydrodynamic model starts decreasing gradually. We associate this with the decrease in energy generation compared to the MHD model seen after $\mathord\approx220\,\mathrm{s}$ in Figure \ref{fig:EnergyGeneration}, which would have the effect of decreasing the convective velocity in the carbon shell. The opposite trend is seen for the MHD Rossby number due to the corresponding increase in energy generation.

At first glance, we find a result that appears in opposition to what is found in solar differential rotation simulations
and the shell burning simulation of \citet{McNeill2022} with a very similar setup. As summarised in \citet{Brun2017}, retrograde rotation is sometimes seen when buoyancy forces dominate the flow (i.e.\ $\mathrm{Ro}\geq 1$), where solar-like stars develop retrograde rotation at the equators and faster rotation at the poles (anti-solar rotation), and faster rotation at the equator for $\mathrm{Ro}\leq 1$. Figure~\ref{fig:Rossby} shows that although we develop strong retrograde rotation in the hydrodynamic model, the Coriolis force dominates its flow in each shell (i.e. $\mathrm{Ro}\leq 1$).

We note, however, a key difference between these models. The anti-solar rotation profiles in low-mass stars develop throughout the convective zone, whereas we find the retrograde motion to be largely concentrated at the base of the burning shells. The different phenomenology could be explained by the rather disparate conditions in surface convection zones in solar-like stars and convection during advanced burning stages. Solar-like stars have an inner radiative zone and a single convective shell above it, and radiation diffusion plays a role both for the internal structure of the convection zone and especially for the structure of the convective boundaries. Our progenitor has three interacting convective shells, radiative effects are unimportant (and not included in the model), and the structure of the boundary is determined by turbulent entrainment.

We suspect that the cause of the retrograde rotation in our simulation is similar to what is described in \citet{Aurnou2007} from hydrodynamic simulations of ice giants, and for solar-like simulations in \citet{Camisassa2022}, where convective rolls exhibit a preferred ``tilt'' in the positive $\phi$ direction. 
The tilted flow structures create a correlation between flows moving inward (outward) and those moving in the negative (positive) $\phi$ direction.
Due to the strong turbulent mixing between shells, however, it is difficult to see in our models whether this tilted flow structure truly arises. For low Rossby convection in the solar case, this usually results in net angular momentum transport away from the rotation axis, which tends to speed up the equator.

\begin{figure}
    \centering
    \includegraphics[width=\columnwidth]{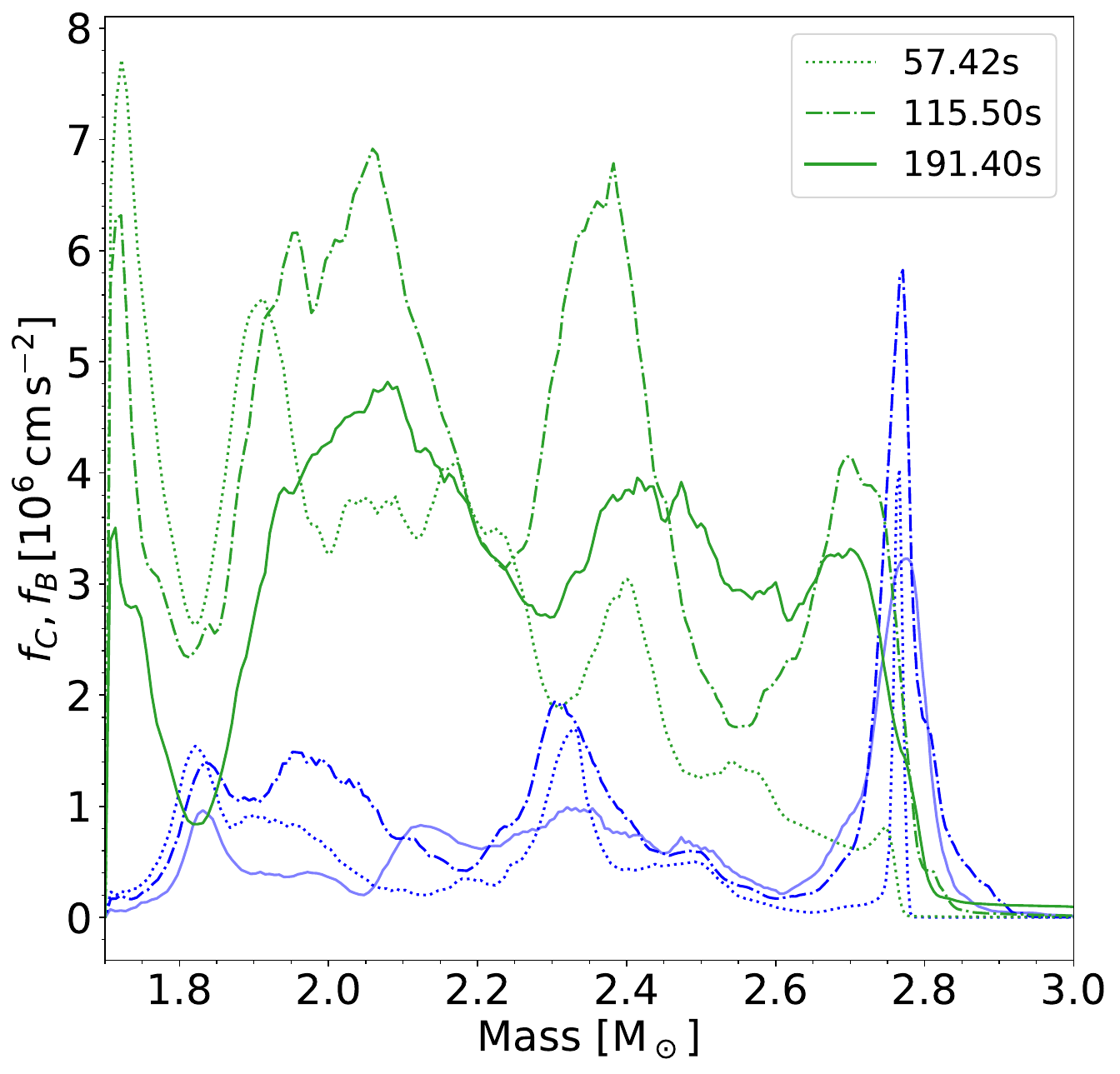}
    \caption{\normalsize Profiles of the Coriolis force (green) and buoyancy force (blue) at $57\,\mathrm{s}$, $115\,\mathrm{s}$, and $191\,\mathrm{s}$ for the purely hydrodynamic model.}
    \label{fig:Forces}
\end{figure}

Since the usual Rossby number characterises the flow in the convection zone globally, it alone would not give us insight into dynamics at the convective boundaries or between convective shells. To understand the interplay between the Coriolis and buoyancy forces at the convective boundaries, we instead plot the angle-averaged magnitude of the buoyancy and Coriolis forces ($f_\mathrm{B}$ and $f_\mathrm{C}$) per unit mass of the hydrodynamic model in Figure~\ref{fig:Forces}. 

\begin{equation}
    f_\mathrm{B} = g\frac{\delta \rho}{\rho_0 }
\end{equation}

\begin{equation}
    f_\mathrm{C} = 2|\boldsymbol{\Omega}\times \mathbf{v}|
\end{equation}

Here $g$ is the gravitational acceleration, $\hat{\rho}$ the RMS averaged density, $\delta \rho$ is the RMS averaged fluctuation from the average density, $\boldsymbol{\Omega}$ is the angular velocity vector of the rotation and $\mathbf{v}$ is the velocity. For simplicity, we assume rotation is confined to $v_\phi$ as in our initial conditions, i.e., $\boldsymbol\Omega$ points in the $z$-direction, giving us only a component for $f_\mathrm{C}$ pointing away from the rotation axis. We plot the RMS average of the absolute value of $f_\mathrm{C}$ in Figure~\ref{fig:Forces}. This is to allow for greater clarity in comparing the two forces, since the retrograde rotation would lead to regions of negative $f_C$. We plot these forces at 57\,s, 115\,s and 190\,s (dotted, dashed and solid lines, respectively). These times were chosen to represent approximate times before the development of retrograde rotation in the hydrodynamic model, when retrograde motion initially begins at the base of the carbon shell, and when it begins at the base of the oxygen shell.

For our simulation, we find the convective boundary overshooting at lower convective boundaries leads to buoyancy forces dominating the Coriolis force in between convective shells ($\mathord\approx 1.85\msun$ and $2.8\msun$, see Figure~\ref{fig:Forces}), which is likely why retrograde motion in our models are confined to the shell boundaries. This is further supported by the fact that we see the ``stable'' retrograde shell start to form at the base of the oxygen shell ($\mathord\approx 1.85\msun$) when the buoyancy force surpasses the Coriolis force. The evolution of the Coriolis force in Figure \ref{fig:Forces} shows that this effect develops due to the transport of angular momentum away from convective boundaries into the convective shells. The magnetic model initially shows a similar force ratio, however, the ratio of buoyancy to Coriolis forces is soon reduced by the rapid rise of magnetic field strength and the subsequent suppression of convection, and hence buoyancy force.

% ----------------------------------------------------------------------------------------

\section{Conclusions}
\label{sec:conclusion} 
We investigated the evolution of magnetic fields
during advanced convective burning stages in massive stars and their backreaction on the flow in a simulation of the oxygen, neon and carbon shells in a rapidly rotating 16$\msun$ progenitor shortly before core collapse. For comparison, we conducted a purely hydrodynamic simulation of the same progenitor as well.
The simulations were run for about 8 minutes of physical time, corresponding to about 32 convective turnovers in the oxygen shell. 

Rapid differential rotation and convection initially amplify the magnetic fields exponentially via the $\alpha\Omega$-dynamo. However, strong magnetic stresses eventually dominate the radial kinetic stresses. The backreaction of the fields on the flow stops the exponential growth
and suppresses convection in the oxygen and neon burning shells. These shells are effectively turned into convectively stable shells by the strong magnetic stresses and continue to burn fuel at the base of the shells without mixing of fuel and ashes. The magnetic field reaches saturation in the oxygen and neon shells after $180\,\mathrm{s}$ (corresponding to $\mathord\approx$12 and 7 convective turnovers respectively). It peaks at $10^{11}\,\mathrm{G}$ in the  oxygen shell but decays to $3\times10^{10}\,\mathrm{G}$ by the end of the simulation. In the carbon shell, the field appears to saturate at $10^{10}\,\mathrm{G}$, but this shell has only completed about one turnover during our entire simulation, so a steady state likely has not been reached. 

The strong magnetic fields that develop also transport angular momentum much more efficiently than in the purely hydrodynamic model.
Already within the short duration of this simulation, the structure transitions from strong differential rotation into a nearly uniform rotation profile, with significant spin-down of the inner shells.

In the purely hydrodynamic model shell convection is sustained and strong differential rotation is maintained. However, it develops a much more complicated rotation profile than in the underlying
1.5D stellar evolution models.  

The emerging rotation profile shows sharp drops at convective boundaries and, during some phases, even shells with retrograde rotation.
We hypothesise that this is due to an instability that occurs when rapid changes of the rotational and convective velocities occur at the convective boundaries, coupled with strong meridional flows towards the poles. The regions with retrograde
rotation are conspicuously associated with spikes in the local Rossby number, i.e., the ratio of the RMS-averaged buoyancy and Coriolis force.
To better understand this phenomenon, we require further studies with different progenitors, varying Rossby number flows, and different grid geometries.  

The transition of the oxygen and neon shells to slowly and rigidly rotation, non-convective region significantly reduces turbulent mixing. While the hydrodynamic model rapidly mixes new material into regions that are burning, the MHD model exhibits sharp drops in oxygen and neon mass fractions in the narrow burning regions. One consequence of this difference is that the hydrodynamic model entrains material deeper into the star, moving the peak of the energy generation of the oxygen shell radially inwards, while the location of peak energy generation of the same shell in the MHD simulation moves outwards. 
Due to the strong temperature sensitivity of oxygen burning ($\propto T^{33}$), this small change in shell position leads to a noticeable change in energy generation between the two models, resulting in increasing nuclear energy generation in the hydrodynamic model and inhibition of nuclear energy generation in the MHD model at late times.

Our results have important implications for core-collapse supernova modelling. For this particular rotating progenitor model, we predict pre-collapse fields of $\mathord\approx2\times 10^{10}\,\mathrm{G}$ in the oxygen shell, similar to what we find for the non-rotating case in \citet{VarmaMuller2021}. Our rotating model exhibits a more gradual drop in field strength with radius. With relatively strong seed fields, we expect less of a delay until magnetic fields can contribute to become relevant for shock revival, i.e.,
by providing an additional ``boost'' to neutrino heating, as seen in \citet{MullerVarma2020, Varma2022}. Due to the suppressed convective flows, the perturbation-aided mechanism \citep{Couch2013, Muller2015a} may be less effective, however, asymmetries seeded by the strong magnetic fields may be enough to deliver a similar effect \citep{Varma2022}. Perhaps most importantly, 
the very rapid redistribution of angular momentum transport from the inner shells casts doubt on the viability of a fast magnetorotational explosion powered by a ``millisecond magnetar''.
For the right conditions to develop, a mechanism would be required to spin up the proto-neutron star during or after the core collapse for a magnetorotational explosion to be launched.
However, there is still work to be done until the findings from simulations of magnetoconvection in rotating stars can be incorporated into models of magnetically- or magnetorotationally-driven explosions. For example, future simulations will need to include the core and self-consistently follow its contraction and incipient collapse to provide initial conditions for supernova simulations.

However, multi-D simulations of rotating massive stars face a much more fundamental challenge.
The MHD model, and to some extent the hydrodynamic model, rapidly diverges from the initial structure of the stellar evolution model. Current stellar evolution models are clearly far from the actual quasi-steady state conditions that would emerge under the influence of rotation, convection and magnetic fields. Ideally, 3D simulations should cover significantly longer time scales to follow the relaxation of the structure into equilibrium and then study the subsequent evolution on secular time scales, but this is clearly beyond current computational resources. It is therefore very important to make 1D stellar evolution models and MHD models more consistent with each other to minimise deleterious effects from big initial transients that limit the fidelity of 3D simulations. This will require improved formalisms for stellar evolution with rotation and magnetic fields \citep{takahashi_20}.
Developing the appropriate methodology for solving the problem of stellar evolution with rotation and magnetic fields by a combination of 1D and 3D modelling is bound to remain an extraordinary and exciting challenge.

\section*{Acknowledgements}
We acknowledge fruitful discussions with R.~Hirschi and A.~Heger. VV acknowledges support from the STFC (Science and Technology Facilities Council; ST/V000543/1).
BM was supported by ARC Future Fellowship FT160100035.  We acknowledge computer time allocations from Astronomy Australia Limited's ASTAC scheme, the National Computational Merit Allocation Scheme (NCMAS), and
from an Australasian Leadership Computing Grant.
Some of this work was performed on the Gadi supercomputer with the assistance of resources and services from the National Computational Infrastructure (NCI), which is supported by the Australian Government, and through support by an Australasian Leadership Computing Grant.  Some of this work was performed on the OzSTAR national facility at Swinburne University of Technology.  OzSTAR is funded by Swinburne University of Technology and the National Collaborative Research Infrastructure Strategy (NCRIS).

%%%%%%%%%%%%%%%%%%%%%%%%%%%%%%%%%%%%%%%%%%%%%%%%%%
\section*{Data Availability}

The data underlying this article will be shared on reasonable request
to the authors, subject to considerations of intellectual property law.



\bibliographystyle{mnras}
\bibliography{paper} % if your bibtex file is called example.bib

%%%%%%%%%%%%%%%%%%%%%%%%%%%%%%%%%%%%%%%%%%%%%%%%%%

%%%%%%%%%%%%%%%%% APPENDICES %%%%%%%%%%%%%%%%%%%%%

% \appendix

% \section{Some extra material}

% If you want to present additional material which would interrupt the flow of the main paper,
% it can be placed in an Appendix which appears after the list of references.

%%%%%%%%%%%%%%%%%%%%%%%%%%%%%%%%%%%%%%%%%%%%%%%%%%

% Don't change these lines
\bsp	% typesetting comment
\label{lastpage}
\end{document}